\documentclass[aps,pra,twocolumn,showpacs,superscriptaddress,floatfix]{revtex4-1}
\usepackage{array}
\usepackage{graphicx,amsmath,amssymb}
\usepackage[usenames]{color}
\usepackage[dvipsnames]{xcolor}
\usepackage[colorlinks=true,linkcolor=blue,anchorcolor=blue,citecolor=blue,urlcolor=blue]{hyperref}
\begin{document}


\title{Two-photon blockade without cascade channel in strong light-matter coupling }

\author{Hang-Hang Han}
\affiliation{School of Physical Science and Technology, Lanzhou University, Lanzhou 730000, China}
\affiliation{Key Laboratory for Quantum Theory and Applications of MoE, Lanzhou Center for Theoretical Physics, Lanzhou University, Lanzhou 730000, China}

\author{Zu-Jian Ying}
\email{Corresponding author: yingzj@lzu.edu.cn}
\affiliation{School of Physical Science and Technology, Lanzhou University, Lanzhou 730000, China}
\affiliation{Key Laboratory for Quantum Theory and Applications of MoE, Lanzhou Center for Theoretical Physics, Lanzhou University, Lanzhou 730000, China}

\begin{abstract}
We report a distinct two-photon blockade (termed as {\sl weak TPB}) discovered in the strong-coupling regime of a parity-broken generalized quantum Rabi model, in addition to the other two-photon blockade (termed as {\sl strong TPB}) in the ultrastrong coupling (USC) regime. In contrast to the strong TPB conventionally with opened cascade decay channel, the weak TPB phase emerges unconventionally with closed cascade decay channel. The closure of cascade channel originally allowed to open by parity breaking is unexpected and its opening in the strong-TPB regime is actually delayed. We extract a cubic law for the cascade transition rate that accounts for the suppression of cascade channel and the delayed opening in the two TPB regimes. Furthermore, we find that the population on the cascade state, which is crucial in the strong TPB, contrarily plays a minor role in the weak TPB. Instead, it is the upper state above the cascade that is relevant for the weak TPB. We demonstrate that the interplay of weak anharmonicity and resonant driving is the primary mechanism responsible for the upper-state population and the formation of weak TPB. Our analyses not only provide deeper insights into the nature of different TPBs, but also imply mechanism and manipulation diversities for the multi-photon blockade, which may open more avenues for developing quantum technologies on the manipulation level of individual quanta.
\end{abstract}

\maketitle

\section{Introduction}

With both the experimental progresses~\cite{Diaz2019RevModPhy,Kockum2019NRP,PRX-Xie-Anistropy,
Qin-ExpLightMatter-2018,WangYouJQ2023DeepStrong,Qin2024PhysRep,LiPengBo-Magnon-PRL-2024} and the theoretical efforts~\cite{PRX-Xie-Anistropy,Braak2011,Boite2020}, light-matter coupling systems~\cite{Eckle-Book-Models,JC-Larson2021} have become an ideal platform and test ground for the explorations of advanced quantum technologies and theories, with the advantages of high controllability and tunability.  Light-matter interactions also have a wide relevance to quantum information and quantum
computation~\cite{Diaz2019RevModPhy,Romero2012,Stassi2020QuComput,Stassi2018,Macri2018},
quantum metrology~\cite{Garbe2020,Montenegro2021-Metrology,Chu2021-Metrology,Garbe2021-Metrology,Ilias2022-Metrology, Ying2022-Metrology,Gietka2023PRL-Squeezing,YangZheng2023SciChina,Hotter2024-Metrology,Alushi2024PRL,Mukhopadhyay2024PRL,Mihailescuy2024,
Ying-Topo-JC-nonHermitian-Fisher,*Ying-Topo-JC-nonHermitian-Fisher-Cover,Ying-g2hz-QFI-2024,*Ying-g2hz-QFI-2024-Cover,
Ying-g1g2hz-QFI-2025,Ying2025g2A4,*Ying2025g2A4-Cover,Ying-g2Stark-QFI-2025,Gietka2025PRL100802,Mihailescu2025CQMtutorial,QiaoFeng2026SpinSqueeze,QiuYi2025gA2},
condensed matter~\cite{Kockum2019NRP}, nanowires~\cite{Nagasawa2013Rings,Ying2016Ellipse,Ying2017curvedSC,Ying2020PRR,Gentile2022NatElec}
and cold atoms~\cite{LinRashbaBECExp2013Review,LinRashbaBECExp2011,LiuYing02025exoticSOC2Ring,*LiuYing02025exoticSOC2Ring-Cover,LiuYing02025KaleidoscopeDDI}.

Indeed, light-matter coupling has opened a rich phenomenology and lead to numerous findings, including integrability~\cite{Braak2011},
finite-component QPTs~\cite{Ashhab2013,Ying2015,Liu2021AQT,Hwang2015PRL,Hwang2016PRL,Irish2017,
Ying-g2hz-QFI-2024,*Ying-g2hz-QFI-2024-Cover,Ying-g1g2hz-QFI-2025,Ying2025g2A4,*Ying2025g2A4-Cover,Ying-g2Stark-QFI-2025,
LiuM2017PRL,Ying-2018-arxiv,Ying2020-nonlinear-bias,Ying-2021-AQT,*Ying-2021-AQT-Cover,
Ying-gapped-top,
Ying-Stark-top,*Ying-Stark-top-Cover,
Ying-Spin-Winding,*Ying-Spin-Winding-Cover,
Ying-JCwinding,Ying-Topo-JC-nonHermitian,*Ying-Topo-JC-nonHermitian-Cover,Ying-Topo-JC-nonHermitian-Fisher,*Ying-Topo-JC-nonHermitian-Fisher-Cover,Ying-gC-by-QFI-2024, Grimaudo2022q2QPT,Grimaudo2023-Entropy,Grimaudo2024PRR,Zhu2024PRL,DeepStrong-JC-Huang-2024,PengJie2019,Padilla2022,Gao2022Rabi-dimer,GaoXL2025SPT},
multicriticalities and multiple points~\cite{Ying2020-nonlinear-bias,Ying-2021-AQT,Ying-gapped-top,Ying-Stark-top}, hidden
symmetry~\cite{Braak2019Symmetry,HiddenSymMangazeev2021,HiddenSymLi2021,HiddenSymBustos2021},
various patterns of symmetry breaking~\cite{Ying2020-nonlinear-bias,Ying-2018-arxiv,Ying-2021-AQT},
universality classification~\cite{Hwang2015PRL,LiuM2017PRL,Irish2017,Ying-2021-AQT,Ying-Stark-top,Ying-Topo-JC-nonHermitian-Fisher,*Ying-Topo-JC-nonHermitian-Fisher-Cover},
spectral collapse~\cite{Felicetti2015-TwoPhotonProcess,e-collpase-Garbe-2017,e-collpase-Duan-2016,CongLei2019,Rico2020} and stablization~\cite{Ying2025g2A4,*Ying2025g2A4-Cover},
spectral conical intersections~\cite{Li2020conical},
classical-quantum correspondence~\cite{Irish-class-quan-corresp},
single-qubit conventional and unconventional topological phase transitions~\cite{Ying-2021-AQT,Ying-gapped-top,Ying-Stark-top,Ying-Spin-Winding,Ying-JCwinding},
coexistence and simultaneous occurrence of Landau-class and topological-class phase transitions~\cite{Ying-2021-AQT,Ying-Stark-top,Ying-JCwinding,Ying-Topo-JC-nonHermitian-Fisher,*Ying-Topo-JC-nonHermitian-Fisher-Cover}, robust topological feature against nonhermiticity~\cite{Ying-Topo-JC-nonHermitian,*Ying-Topo-JC-nonHermitian-Cover,Ying-Topo-JC-nonHermitian-Fisher,*Ying-Topo-JC-nonHermitian-Fisher-Cover},
squeezing and critical resources for quantum metrology~\cite{Garbe2020,Montenegro2021-Metrology,Chu2021-Metrology,Garbe2021-Metrology,Ilias2022-Metrology, Ying2022-Metrology,YangZheng2023SciChina,Gietka2023PRL-Squeezing,Hotter2024-Metrology,Alushi2024PRL,Mukhopadhyay2024PRL,Mihailescuy2024,
Ying-Topo-JC-nonHermitian-Fisher,*Ying-Topo-JC-nonHermitian-Fisher-Cover,Ying-g2hz-QFI-2024,*Ying-g2hz-QFI-2024-Cover,
Ying-g1g2hz-QFI-2025,Ying2025g2A4,*Ying2025g2A4-Cover,Ying-g2Stark-QFI-2025,Gietka2025PRL100802,Mihailescu2025CQMtutorial,QiuYi2025gA2},
photon blockade effect~\cite{Boite2016-Photon-Blockade,Ridolfo2012-Photon-Blockade,LiaoJQ2020BlockadeJC,Ma2026PRL-Strong-2PB,Felicetti2026PRXQuantumBlockade,Carmichael1985,Birnbaum2005,Shamailov2010,Liew2010,Hamsen2017,Garziano2017,Flayac2017},
and so forth. Among these achievements, one particularly interesting phenomenon emerging from input-output theories and measurements is the photon blockade.
Photon blockade~\cite{Boite2016-Photon-Blockade,Ridolfo2012-Photon-Blockade,LiaoJQ2020BlockadeJC,Ma2026PRL-Strong-2PB,Felicetti2026PRXQuantumBlockade,Carmichael1985,Birnbaum2005,Shamailov2010,Liew2010,Hamsen2017,Garziano2017,Flayac2017}, as an analogue to Coulomb blockade, is a fundamental quantum optical phenomenon in which the absorption of one photon strongly suppresses the absorption of subsequent photons, leading to antibunched photon statistics. Photon blockade is one of the clearest demonstrations that light can be manipulated at the level of individual quanta. In cavity quantum electrodynamics (QED) and circuit QED systems, this effect
acts both as a probe of strong light-matter interaction and as a resource for generating non-classical states of light.

In the Jaynes-Cummings (JC) regime~\cite{JC-model,JC-Larson2021} where counter-rotating terms are negligible, conventional one-photon blockade was first theoretically analyzed in the context of photon antibunching for a single atom in a resonant cavity~\cite{Carmichael1985} and later experimentally demonstrated in an optical cavity with one trapped atom~\cite{Birnbaum2005}, as characterized by $g^{(n)}(0)<1$ where $g^{(n)}(0)$ ($n \geqslant 2$) are the $n$th-order equal-time correlation functions. This conventional blockade arises from the inherent anharmonicity of the dressed-state ladder. Multi-photon blockade phenomena in JC-type systems have also been extensively studied, including two-photon blockade~\cite{Hamsen2017} and higher-order multiphoton blockade~\cite{LiaoJQ2020BlockadeJC,Shamailov2010}. In parallel, the unconventional photon blockade mechanism based on quantum interference was proposed and developed in coupled quantum systems~\cite{Liew2010,Flayac2017}.

In the ultrastrong-coupling (USC) regime~\cite{Diaz2019RevModPhy,Kockum2019NRP,Qin2024PhysRep}, where the light-matter coupling strength $g$ becomes comparable to the bare frequency $\omega_0$, counter-rotating terms become significant~\cite{PRX-Xie-Anistropy} and the physics is substantially enriched. The standard photon blockade scenario is significantly modified, with parametric processes playing an important role~\cite{Ridolfo2012-Photon-Blockade}. Photon bunching from individual dressed qubits has also been predicted in this regime~\cite{Garziano2017}. A prominent example is the strong two-photon (or two-polariton) blockade (TPB), in which two-photon bunching ($g^{(2)}(0) > 1 $) and suppression of higher-order clustering ($g^{(3)}(0) < 1 $) appear when the cascade decay channel from the second excited state $|\psi_{1+}\rangle $ to the first excited state $|\psi_{1-}\rangle $ opens at sufficiently large $g$~\cite{Ma2026PRL-Strong-2PB}. This mechanism relies on the population of the lower photon state $|\psi_{1-}\rangle $ via real cascade transitions, followed by destructive interference that suppresses three-photon processes while allowing two-photon bunching. The fate of photon blockade in the deep strong-coupling regime has also been theoretically investigated~\cite{Boite2016-Photon-Blockade}.

Comprehensive reviews of ultrastrong light-matter coupling, including its impact on photon blockade, are available in the literature~\cite{Kockum2019NRP,Diaz2019RevModPhy}. However, the intermediate coupling regime---where $g$ is large enough for counter-rotating terms and parity-breaking effects to matter, yet still below the threshold for significant cascade-channel opening---has been less explored.  In particular, it is not obvious whether TPB can emerge through mechanisms other than cascade decay, and it is not clear how parity breaking modifies the selection rules and interference pathways that govern multi-photon statistics.

In this work, we address these issues by studying a parity-broken generalized quantum Rabi model with coherent-field driving. Apart from a TPB in the USC regime which we label by {\sl strong TPB}, we discover another distinct TPB (labeled as {\sl weak TPB}) occurring at intermediate coupling strengths ($0.037 \lesssim g/\omega_0 \lesssim 0.043$) which belongs to conventional strong-coupling regime. We find that the weak TPB manifests different characters and is essentially distinguished from the strong TPB in origins. The findings about these different TPBs include:
(i) While the strong TPB occurs conventionally with opening of the key cascade decay channel from $|\psi_{1+}\rangle$ to $ |\psi_{1-}\rangle $, in the cascade series $|\psi_{1+}\rangle \rightarrow |\psi_{1-}\rangle \rightarrow |\psi_0\rangle$, the weak TPB appears unconventionally with the cascade channel closed.
(ii) The cascade channel originally allowed to open by broken parity symmetry is unexpectedly suppressed in the weak TPB regime and its opening in the strong TPB is actually delayed. We extract a cubic law in the cascade transition rate which accounts for its unexpected suppression and delayed opening in the two TPB regimes.
(iii) While the population of the cascade middle state is crucial in the strong TPB, it contrarily plays a minor
role in the weak TPB. Instead, it is the upper state $|\psi_{2+}\rangle$ above the cascade that is crucial for the weak TPB.
(iv) In contrast to the mechanism of the cascade decay in dissipation for the strong TPB, the weak anharmonicity in driving is the primary origin for the weak TPB. The weak anharmonicity leads to different populations of the upper state $|\psi_{2+}\rangle$ in the resonant driving, which qualitatively changes the correlation functions $g^{(2)}(0)$ and $g^{(3)}(0)$ and gives rise to the formation of the weak TPB.

Our findings reveal that TPB is not merely a phenomenon tied exclusively to cascade decay, but can arise from qualitatively different microscopic mechanisms depending on the coupling regime. Our mechanism tracking also implies that manipulation of the TPBs may be intervened in different stages, as in the processes of dissipation and driving here.  These insights may help to open novel avenues for engineering non-classical light statistics on the manipulation level of individual quanta, which can be realized in superconducting circuit-QED systems where both coupling strength and parity symmetry is highly tunable and controllable.

This paper is organized as follows. Section~\ref{Section-Model} introduces
parity-broken generalized quantum Rabi model driven
resonantly into the upper photon.
Section~\ref{Section-input-output} formulates the master equation in dissipation and introduces input-output quantities in the dressed picture.
Section~\ref{Section-TPB} reveals the weak TPB in the intermediate coupling regime, in addition to the strong TPB in the USC regime.
Sections~\ref{Sect-Mechanism-Strong-TPB} and~\ref{Sect-weak-TPB-mechanism} clarify the mechanisms for the strong TPB and weak TPB, respectively.
In Sec.~\ref{Sect-Compare}, a tabular comparison is given for the differences of the strong TPB and weak TPB.
Finally, Section~\ref{Section-Conclusion} summarizes our conclusions.

\section{Model }\label{Section-Model}

We consider the generalized quantum Rabi Hamiltonian describing a flux qubit coupled to a transmission-line resonator,
\begin{equation}
H_0 = \omega_c a^\dagger a + \omega_g \sigma^+ \sigma^- + g (a + a^\dagger) \bigl( \cos\theta \, \sigma_z - \sin\theta \, \sigma_x \bigr),
\label{eq:H0}
\end{equation}
where $a$ ($a^\dagger$) is the annihilation (creation) operator of the resonator mode, $\sigma^\pm$ and $\sigma^{x,z}$ are Pauli operators of the qubit, and $\theta$ is the coupling phase controlled by the external flux bias. When $\theta = m\pi/2$, the Hamiltonian possesses the parity symmetry as it commutes with the parity operator $\hat{P}=\sigma_z \exp(i \pi a^\dagger a)$. For the value $\theta = 0.3\pi$ used in this work, the parity symmetry is broken, allowing dissipation decay that was previously forbidden in parity symmetry to acquire non-zero matrix elements. The system is driven by a weak coherent field,
\begin{equation}
H_{\rm drive} = \Omega \cos(\omega_l t)(a + a^\dagger),
\end{equation}
with drive amplitude $\Omega$ and frequency $\omega_l$. The total Hamiltonian is $H = H_0 + H_{\rm drive}$. Throughout this work we set $\omega_g = \omega_c \equiv \omega_0$ and $\hbar = 1$ as the units.

\begin{figure}[thbp]
	\centering
	\includegraphics[width=0.88\linewidth]{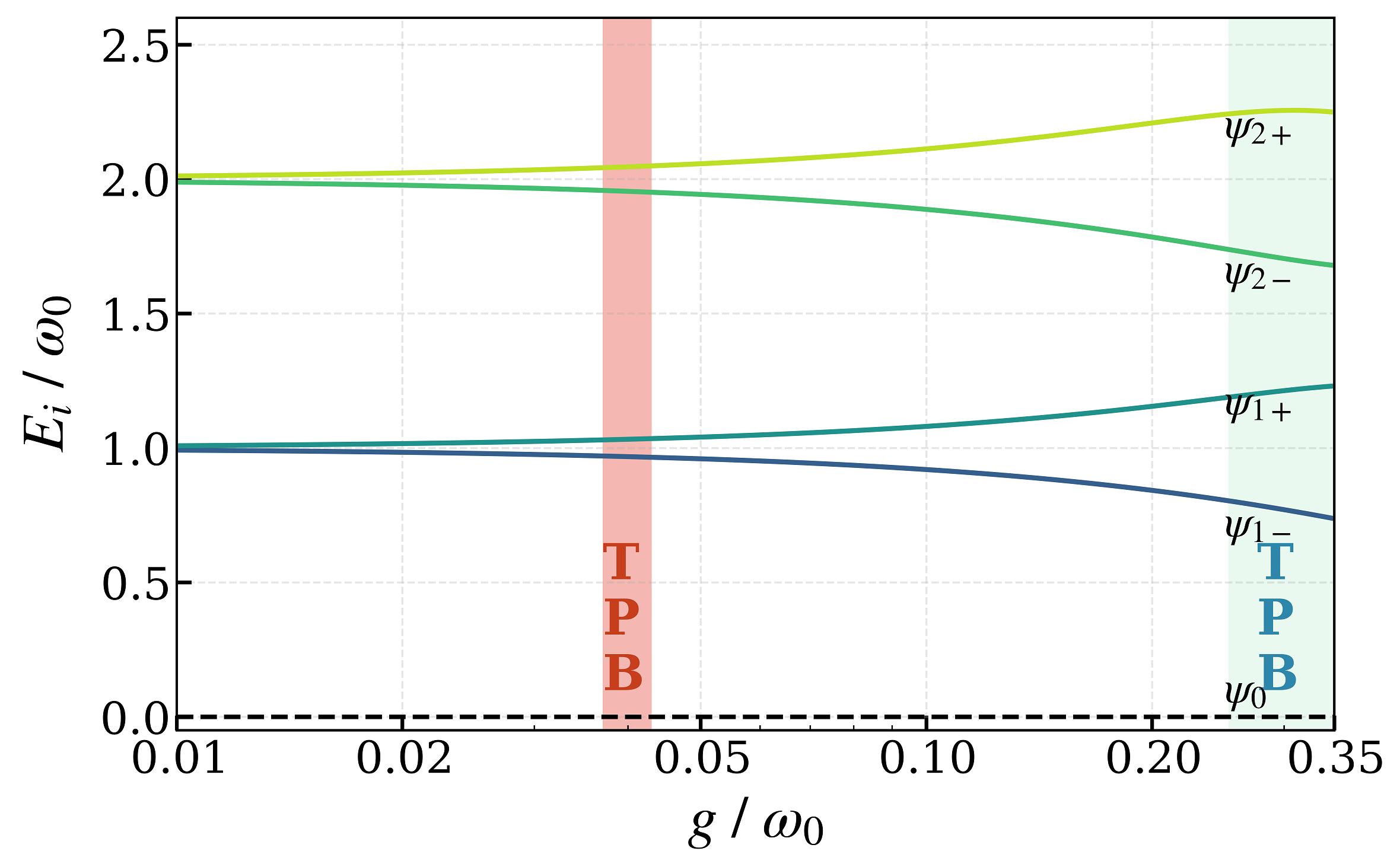}
	\caption{Relevant low energy levels $E_i$ of $H_0$ as functions of coupling strength $g$. Here, $\omega_g = \omega_c \equiv \omega_0$ and $\theta = 0.3\pi $, and the energy of the ground state $|\psi_0\rangle$ is set as the energy reference. The pale-rose (gray) shaded region marks the {\sl weak-TPB} phase and the light-minty-green (light gray) shaded region denotes the {\sl strong-TPB} regime, as revealed in Fig.~\ref{fig:g2g3}.}
\label{fig-Ej}
\end{figure}

In Fig.\ref{fig-Ej} we show some low energy levels of the system Hamiltonian $H_0$, including the ground state $|\psi_0\rangle$ and two splitting sets of photon states. For the study of TPB in the present work, the system will be resonantly driven from the ground state $|\psi_0\rangle$ onto the second excited state $|\psi_{1+}\rangle$ by tuning the driving frequency to be the excitation energy of $|\psi_{1+}\rangle$.

\section{Master equation and input-output quantities}\label{Section-input-output}

In this work, the coupling strength is varied from $  g/\omega_0 = 0.01  $ to $  0.35  $. Although the standard  input-output relation remains valid at strong-coupling regime, counter-rotating terms become significant in the ultrastrong-coupling regime. For uniformity, we therefore adopt the modified input-output formalism across the entire parameter range. To correctly describe dissipation, we work entirely in the dressed-state basis obtained from the full $H_0$. The full Hamiltonian $H_0$ (including counter-rotating terms) is exactly diagonalized in a large Fock space to obtain the dressed eigenstates $|\psi_j\rangle$ and eigenenergies $E_j$ satisfying $H_0 |\psi_j\rangle = E_j |\psi_j\rangle$. We label the ground state as $|\psi_0\rangle$, the lower one-photon state as $ |\psi_{1-}\rangle$, and the upper one-photon state as $ |\psi_{1+}\rangle$.
The relevant transition frequencies are denoted $\Delta_{jk} = E_j - E_k$.

The Lindblad master equation takes the form
\begin{equation}
\dot{\rho} = -i [H_{\rm rot}, \rho] + \sum_{c = a,\sigma^-} \mathcal{L}_c[\rho]+\mathcal{L}_{deph}[\rho],
\end{equation}
where $\mathcal{D}[L]\rho  $ is the standard Lindblad dissipator in the dressed basis. The second term represents the dissipator
\begin{equation}
\mathcal{L}_c[\rho] = \sum_{j > k} \Gamma_{jk}^c \mathcal{D}\bigl[|\psi_k\rangle\langle\psi_j|\bigr] \rho,
\end{equation}
with the transition rates
\begin{equation}
\Gamma_{jk}^c = \gamma_c \frac{|\Delta_{jk}|}{\omega_0} |C_{jk}^{(c)}|^2. \label{eq:t10}
\end{equation}
Here, $\gamma_c$ is the bare decay rate of operator $c$ and the matrix elements are defined as
\begin{equation}
C_{jk}^{(c)} = -i \langle \psi_j | (c - c^\dagger) | \psi_k \rangle \quad (c = a, \sigma^-). \label{Eq-Cjk}
\end{equation}
The laster term in the master equation denotes the qubit dephasing
\begin{equation}
\mathcal{L}_{\rm deph}[\rho] = \sum_j \Gamma_j^{\rm deph} \mathcal{D}[|j\rangle\langle j|]\rho,
\end{equation}
where $  \Gamma_j^{\rm deph} = \gamma_{\rm deph} \big|\langle j|\sigma_z|j\rangle\big|  $.

Following the formalism appropriate for all couplings, the output field operator is related to the input field by~\cite{Ridolfo2012-Photon-Blockade}
\begin{equation}
a_{\rm out}(t) = a_{\rm in}(t) - i \Upsilon \dot{X}^+(t), \label{Eq-input-ouput}
\end{equation}
where $\Upsilon$ is the waveguide coupling coefficient and $\dot{X}^+$ is the positive-frequency component of the time derivative of the  operator $X = -i(a -a^\dagger)$, expressed in the dressed-state basis:
\begin{equation}
\dot{X}^+ = -i \sum_{j > k} \Delta_{kj} \, X_{jk} \, |\psi_j\rangle\langle\psi_k|,
\end{equation}
with $X_{jk} = \langle\psi_j | X | \psi_k \rangle$.

The $n$th-order equal-time correlation functions of the output field are then defined as
\begin{equation}
g^{(n)}(0) = \frac{\langle (\dot{X}^-)^n (\dot{X}^+)^n \rangle}{\langle \dot{X}^- \dot{X}^+ \rangle^n},
\label{eq:gn}
\end{equation}
where $\dot{X}^-$ is the Hermitian conjugate of $\dot{X}^+$. This definition properly incorporates both photonic and atomic contributions to the output field statistics in the presence of strong counter-rotating terms and parity breaking.

\begin{figure}[t]
	\centering
	\includegraphics[width=0.9\linewidth]{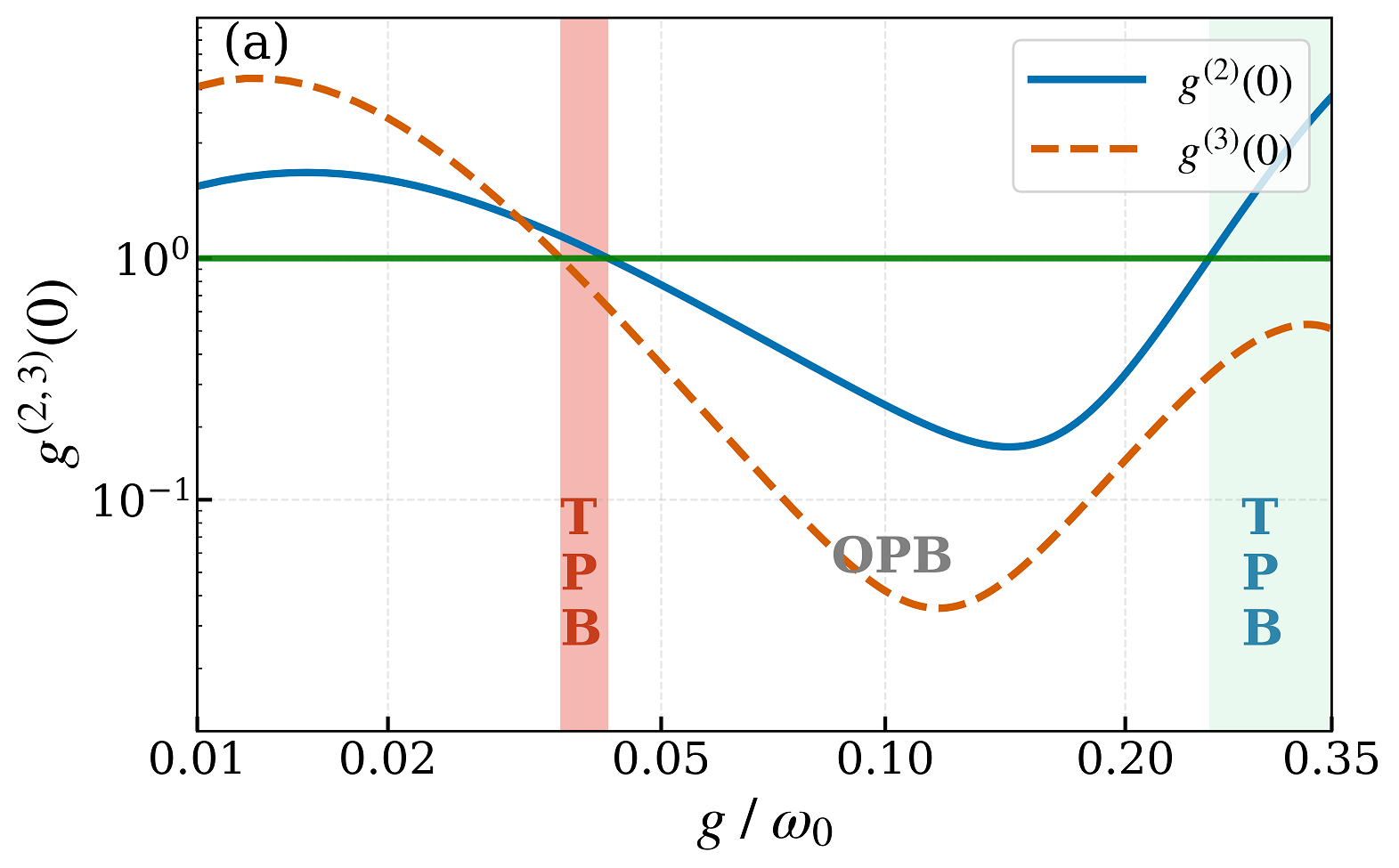}
	\includegraphics[width=0.9\linewidth]{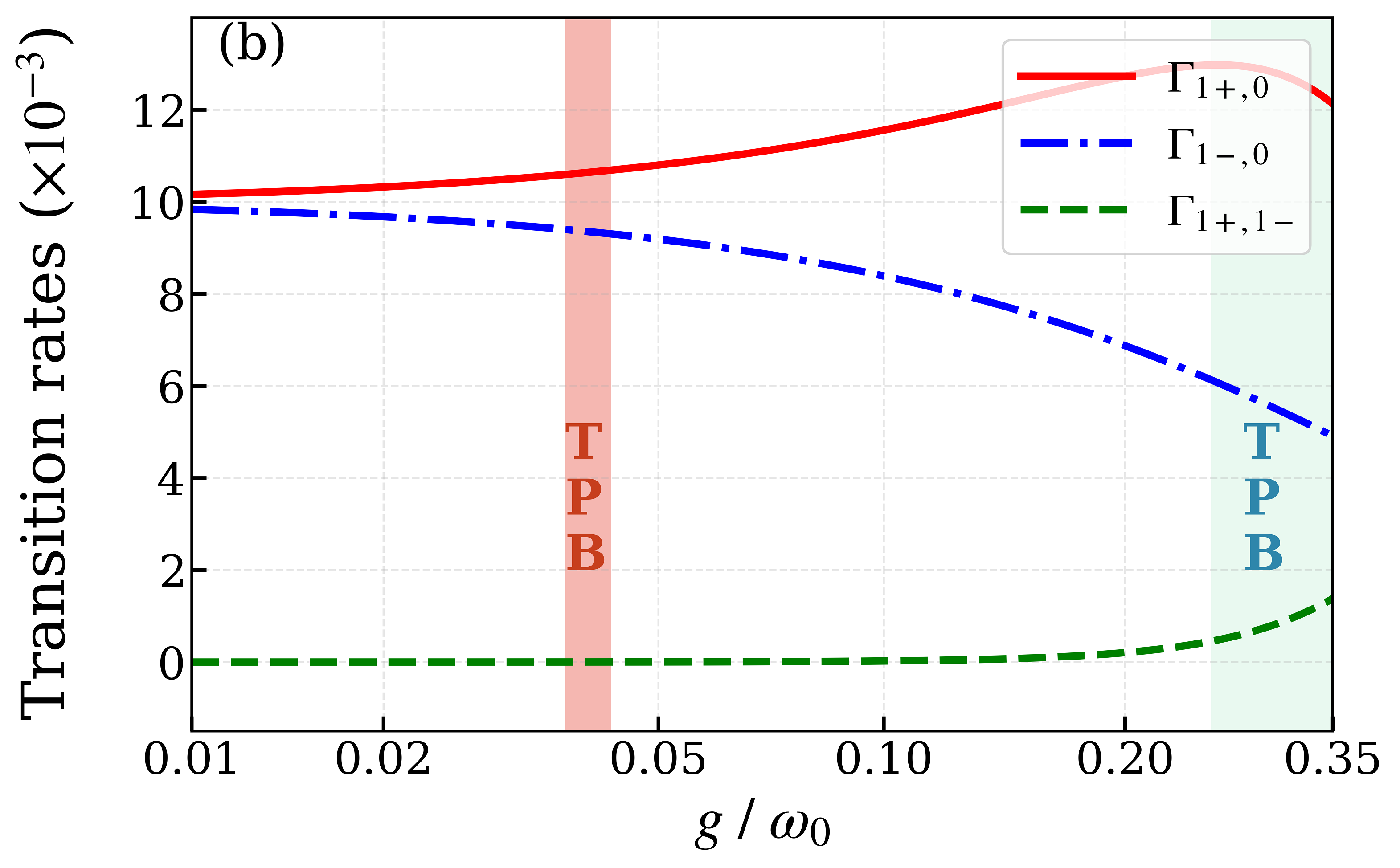}
	\caption{{\it Finding two-photon blockade (TPB) in strong-coupling regime ($0.01<g/\omega_0<0.1$) without cascade channel, apart from the TPB in ultrastrong-coupling (USC, $g/\omega_0>0.1$) regime with cascade channel.} (a) The second- and third-order equal-time correlation functions $g^{(2)}(0)$ (solid) and $g^{(3)}(0)$ (dashed) as functions of the coupling strength $g$ under resonant driving from the ground state onto the $|\psi_{1+}\rangle$ state. The pale-rose (gray) shaded region marks the {\sl weak-TPB} window ($0.037 \le g/\omega_0 \le 0.043$), where $g^{(2)}(0) > 1$ while $g^{(3)}(0) < 1$.
(b) Transition rates versus $g/\omega_0$: $\Gamma_{1+,0}$ (solid red), $\Gamma_{1-,0}$ (dash-dotted blue), and $\Gamma_{1+,1-}$ (dashed green). The cascade channel $\Gamma_{1+,1-}$ remains negligible throughout the weak-TPB regime and becomes significant only near $g \approx 0.253\omega_0$, where the {\sl strong-TPB} regime starts [light-minty-green (light gray) shaded region]. Here $\Omega = 10^{-3}\omega_0 $, $\gamma_a = \gamma_{\sigma^-} =2 \gamma_{\rm deph}=10^{-2}\omega_0  $ and other system parameters are the same as in Fig.~\ref{fig-Ej}. Unless otherwise specified, the parameters are the same in other figures.
}
\label{fig:g2g3}
\end{figure}

\section{Different Two-Photon Blockades}\label{Section-TPB}

To identify possible signatures of TPB, we compute the second- and third-order equal-time correlation functions $g^{(2)}(0)  $ and $g^{(3)}(0)  $, defined in Eq.~(\ref{eq:gn}), as functions of the coupling strength $g$. The system is driven resonantly at the upper-photon frequency $  \omega_d = E_{1+} - E_0  $, with the system parameters
$\theta = 0.3\pi $, $\Omega = 10^{-3}\omega_0  $, $ \gamma_a = \gamma_{\sigma^-} =2 \gamma_{\rm deph}=10^{-2}\omega_0  $~\cite{Ma2026PRL-Strong-2PB}. As mentioned in Introduction, the occurrence of one-photon blockade (OPB) is judged by correlation functions reduced from unity $g^{(2)}(0)<1$ and $g^{(3)}(0)<1$, while the TPB is indentified by $g^{(2)}(0)>1$ and $g^{(3)}(0)<1$~\cite{Boite2016-Photon-Blockade,Ridolfo2012-Photon-Blockade,LiaoJQ2020BlockadeJC,Ma2026PRL-Strong-2PB,Carmichael1985,Birnbaum2005,Shamailov2010,Liew2010,Hamsen2017,Garziano2017,Flayac2017}.

\subsection{Apart from conventional OPB and strong TPB: Emerging of weak TPB}

As shown by the evolutions of correlation functions $g^{(2)}(0) $ and $g^{(3)}(0) $ in Fig.~\ref{fig:g2g3}(a), four distinct regimes emerge in different coupling strengths:

(i) In weak couplings $g/\omega_0 \lesssim 0.037$, both $g^{(2)}(0)$ (solid line) and $g^{(3)}(0)$ (dashed line) exceed unity, which indicates photon bunching. There is no photon blockade in this regime.

(ii) With strong couplings $ 0.043 \lesssim g/\omega_0 \lesssim  0.253$, we have both $g^{(2)}(0)<0$ and $g^{(3)}(0)<0$, which are characters of the conventional one-photon blockade (OPB).

(iii) For larger couplings exceeding $g \simeq 0.253\omega_0$ as in the light-minty-green (light gray) shaded region, which actually belongs to the USC regime ($g \gtrsim 0.1\omega_0$), one sees $g^{(2)}(0) > 1 $ while $g^{(3)}(0) < 1 $, which identifies two-photon blockade (TPB). Here we refer to this regime as {\sl strong TPB} regime, in the sense that not only the coupling is strong in USC regime but also the amplitude of the two-photon correlation function $g^{(3)}(0)$ is reduced from the unity relatively more than the other next revealed weak-TPB regime.

(iv) In strong couplings, we find a window $0.037 \lesssim g/\omega_0 \lesssim 0.043$, as in the pale-rose (gray) shaded region, in which the correlation functions also manifest the TPB characteristics $g^{(2)}(0) > 1 $ and $g^{(3)}(0) < 1 $. We refer to this regime as {\sl weak TPB} regime, in the sense that the coupling is weaker relatively to the strong TPB in the USC regime and also the amplitude of the two-photon correlation function $g^{(3)}(0)$ is reduced from the unity in a degree less than the strong TPB in regime (iii).

Here, some clarifications should be made on the conventional terms for coupling strength. The USC ($0.1<g/\omega_0<1$) and deep-strong coupling ($g/\omega_0>1.0$) are termed with the cavity frequency $\omega_0$ as the comparison reference, while conventional weak coupling ($g/\{\kappa,\gamma\}<1$) and strong coupling ($g/\{\kappa,\gamma\}>1$) differently take the cavity and qubit dacay rates $\{\kappa,\gamma\}$ as the reference~\cite{Kockum2019NRP}. Here, for the sake of reference unification and description convenience, in the above we have not adopted the conventional terms for the weak coupling and strong coupling. Nevertheless, considering the order of the dissipation rate $\gamma \sim 0.01 \omega _0$, the weak-TPB window $0.037 \lesssim g/\omega_0 \lesssim 0.043$, in item (iv) with ``strong couplings", really belongs to the conventional strong-coupling regime.

\subsection{Difference of the weak TPB and strong TPB: Closed cascade channel and opened cascade channel}

A close look at the transition rates in dissipation will reveal the essential difference of the weak TPB and the strong TPB. Figure~\ref{fig:g2g3}(b) shows the transition rates $\Gamma_{1+,0}$ (solid red), $\Gamma_{1-,0}$ (dash-dotted blue), and $\Gamma_{1+,1-}$ (dashed green), which represent the cascade channels after driving from $|\psi_{0}\rangle$ onto $|\psi_{1+,0}\rangle$:
\begin{eqnarray}
\Gamma_{1+,0}:&& \quad |\psi_{1+}\rangle  \rightarrow  |\psi_{0}\rangle, \\
\Gamma_{1-,0}:&& \quad |\psi_{1-}\rangle  \rightarrow  |\psi_{0}\rangle, \\
\Gamma_{1+,1-}:&& \quad |\psi_{1+}\rangle  \rightarrow  |\psi_{1-}\rangle.
\end{eqnarray}
We see that the cascade channels $\Gamma_{1+,0}$ and $\Gamma_{1-,0}$ are always open as indicated by their finite values in the solid and dot-dashed lines in Fig~\ref{fig:g2g3}(b). In this situation the last channel $\Gamma_{1+,1-}$ is particularly important, as its opening will lead to the conventional TPB. Indeed, from the dashed line in Fig~\ref{fig:g2g3}(b), around $g=0.253\omega_0$ we see the appreciable rising of $\Gamma_{1+,1-}$ from the vanishing value, which means the opening of this cascade channel and corresponds the transition from the OPB to the strong TPB. Therefore, the strong TPB is the conventional TPB which originates from the opening of cascade channel.

In contrast, the weak TPB lies in the regime far before the OPB/strong-TPB transition, where $\Gamma_{1+,1-}$ is completely suppressed and consequently the cascade channel $\Gamma_{1+,1-}$ is well closed. Thus, we see the essential difference of the weak TPB from the strong TPB:
\begin{eqnarray}
\text{Strong TPB:}&& \quad \Gamma_{1+,1-} \text{ cascade channel opened}, \\
\text{Weak TPB:}&& \quad \Gamma_{1+,1-} \text{ cascade channel closed},
\end{eqnarray}
which indicates that TPB can arise from distinct physical mechanisms at different coupling strengths, not necessarily relying on the opening of cascade decay.

In the following we will make a detailed analysis of the physical mechanisms for both the strong TPB and the weak TPB.

\section{Physical mechanism for strong TPB: Delayed opening of cascade channel}\label{Sect-Mechanism-Strong-TPB}

Although numerical simulations~\cite{Ma2026PRL-Strong-2PB} reveal that the cascade channel $|\psi_{1+}\rangle\to|\psi_{1-}\rangle$ opens appreciably only when the coupling strength exceeds approximately $0.253\omega_0$, an explanation for the essential origin for the closing and opening of this cascade channel is still lacking. In this section we will clarify the underlying mechanism for the OPB/strong-TPB transition.

\subsection{Unexpected closure of cascade channel $\Gamma_{1+,1-}  $}

Actually the closure of the cascade channel $\Gamma_{1+,1-}$ before the strong-TPB regime is unexpected. If the system had parity symmetry, the cascade channel $\Gamma_{1+,1-}$ would be prohibited as the dissipation-induced transition element $C_{jk}^{(c)}$ is vanishing for states in the same parity. However, for the model $H_0$ considered in the present work, the parity symmetry is explicitly broken. Therefore, one would expect that $\Gamma_{1+,1-}$ is not vanishing and this cascade channel should be open. However, in reality the cascade channel $\Gamma_{1+,1-}$ is contrarily closed from the weak coupling up to USC regime until the transition around $g\simeq 0.253 \omega_0$, as observed from the dashed line in Fig.~\ref{fig:g2g3}(b). There must be some underlying origin for such unexpected channel closure before the transition. We find that the unexpected channel closure originates from a cubic law in the coupling dependence of $\Gamma_{1+,1-}$, which leads to a delayed opening of cascade channel, as uncovered below.

\begin{figure}[t!]
	\centering
	\includegraphics[width=0.99\linewidth]{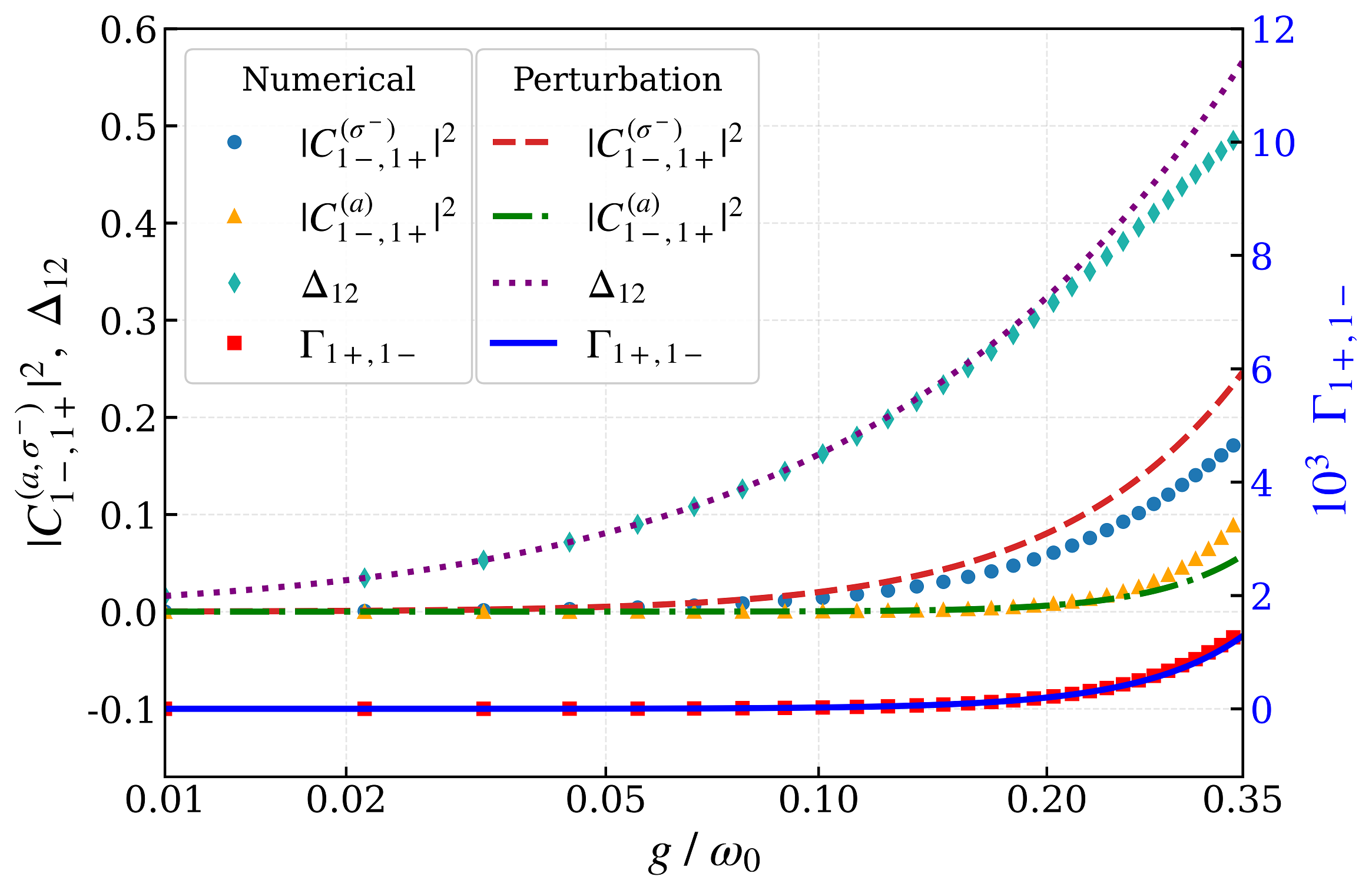}
	\caption{{\it The $g^3$ law of $\Gamma_{1+,1-}$ and delayed opening of the cascade channel.}
The lines represent perturbative results of $|C_{1-,1+}^{(c)}|^2$ with $c=\sigma^-$ (dashed) and $c = a$ (dot-dashed), the energy level splitting $\Delta_{12}=E_{1+}-E_{1-}$ (dotted),  and the final transition rate $\Gamma_{1+,1-}$ (solid), with $\mathcal{O}[(g/\omega_0)^1)]$, $\mathcal{O}[(g/\omega_0)^2)]$, $\mathcal{O}[(g/\omega_0)^1)]$, $\mathcal{O}[(g/\omega_0)^3)]$ as the leading orders, respectively. The symbols denote the corresponding numerical results.
The final transition rate $\Gamma_{1+,1-}$ is strongly suppressed in the weak-TPB regime and the opening of the $\Gamma_{1+,1-}$ channel originally allowed by parity breaking is delayed in the strong-TPB regime, due to the cubic law.
}
\label{fig:annum}
\end{figure}

\subsection{Cubic law in the coupling dependence of $\Gamma_{1+,1-}  $}

To unravel the underlying mechanism of the closure of cascade channel $%
\Gamma _{1+,1-}$ in an analytical way, we employ perturbation theory to the
dressed states. We decompose the system Hamiltonian into the exactly
solvable JC part $H_{\mathrm{JC}}$ and the rest part $V$ as the
perturbation, while $V$ contains all counter-rotating and longitudinal
terms. Then the eigen states $|\psi _{1\pm }^{\left( 0\right) }\rangle $ of $%
H_{\mathrm{JC}}$ within $N=1$, where $N=a^{\dagger }a+\sigma _{z}/2+1/2$ is
the excitation number, are taken as the unperturbed states with unperturbed
energies $E_{1+}^{(0)}$ and $E_{1-}^{(0)}$, respectively. Thus, the
dissipative matrix elements controlling the cascade decay, defined in Eq.~(%
\ref{eq:t10})
\begin{equation}
|C_{1-,1+}^{(c)}|=|\langle \psi _{1-}|(c-c^{\dagger })|\psi _{1+}\rangle
|\quad (c=a,\sigma ^{-}),
\end{equation}%
admits the perturbative expansion
\begin{equation}
|C_{1-,1+}^{(c)}|=|C_{1-,1+}^{(c)}|_{00}+|C_{1-,1+}^{(c)}|_{01}+\mathcal{O}
(g^{3}).
\end{equation}

The zeroth-order term vanishes identically (see Appendix \ref{Appendix-zero-order}),
\begin{equation}
|C_{1-,1+}^{(c)}|_{00}=\left\vert \langle \psi _{1-}|(c-c^{\dagger })|\psi
_{1+}\rangle ^{\left( 0\right) }\right\vert =0,
\end{equation}%
due to that $c$ induces transitions from the $N=1$ sector to bases in other $%
N$ sectors which are orthogonal to the $N=1$ sector. Here, the superscript
(0) denotes the expectation over the unperturbed states.

With the definition $X^{c}=c-c^{\dagger }\quad (c=a,\sigma ^{-})$, the
first-order correction reads
\begin{equation}
\begin{split}
|C_{1-,1+}^{(c)}|_{01}=\sum_{v\notin \{1+,1-\}}& \Biggl[\frac{\langle \psi
_{1-}|X^{c}|v\rangle ^{\left( 0\right) }\langle v|V|\psi _{1+}\rangle
^{\left( 0\right) }}{E_{1+}^{(0)}-E_{v}^{(0)}} \\
& +\frac{\langle \psi _{1-}|V|v\rangle ^{\left( 0\right) }\langle
v|X^{c}|\psi _{1+}\rangle ^{\left( 0\right) }}{E_{1-}^{(0)}-E_{v}^{(0)}}%
\Biggr].
\end{split}
\end{equation}
where $v$ sums over unperturbed states except the $N=1$ sector. A
complete analytical evaluation of the contributions from the relevant
sectors ($N=0$ and $N=2$), including explicit term-by-term sign tracking,
shows that the linear-order part of $|C_{1-,1+}^{(a)}|_{01}$ vanishes exactly
and the quadratic part of $|C_{1-,1+}^{(\sigma ^{-})}|_{01}$ is exactly zero.
Thus, we have the leading orders in expansions
\begin{equation}
\begin{split}
|C_{1-,1+}^{(a)}|& =C_{a}(\theta )\left( \frac{g}{\omega _{0}}\right) ^{2}+%
\mathcal{O}(g^{3}), \\
|C_{1-,1+}^{(\sigma ^{-})}|& =C_{\sigma ^{-}}(\theta )\left( \frac{g}{\omega
_{0}}\right) +\mathcal{O}(g^{3}),
\end{split}
\label{eq-C-Matrix}
\end{equation}%
where $C_{a}(\theta )$ and $C_{\sigma ^{-}}(\theta )$ are constant
coefficients with respect to the variation of coupling. The detailed proof is
given in Appendix~\ref{Appendix-proof-M-g2}.

Note that the transition rate of the cascade channel is proportional to the
square of the dissipative matrix element. Consequently, the transition rate from the contribution of the dissipative matrix element
becomes quadratic in the leading order of coupling dependence.
More explicitly, the transition rate $\Gamma_{1^+,1^-}$ defined in Eq.~(\ref%
{eq:t10}) becomes
\begin{equation}
\Gamma_{1^+,1^-}= \frac{\gamma\Delta_{12}}{\omega_0}\left[%
C_{\sigma^{-}}^2(\theta) \left( \frac{g}{\omega_0} \right)^2+ C_a^2(\theta)
\left( \frac{g}{\omega_0} \right)^4 +\mathcal{O}(g^5) \right],
\end{equation}
where $\gamma= \gamma_a = \gamma_{\sigma^-}$. Furthermore, $%
\Delta_{12}=E_{1+}-E_{1-}$ denotes the energy splitting between the two
states$| \psi_{1-} \rangle$ and $| \psi_{1+} \rangle$, which actually has a linear
and cubic dependence in the first two orders (see the derivation in Appendix~%
\ref{Appendix-gap})
\begin{equation}
\Delta_{12} = 2g\beta - (8 \alpha^2 \beta+\beta^3/2)g^3 + \mathcal{O}(g^4).
\end{equation}
As a result, the transition rate of the cascade channel finally upgrades to
a cubic law in the leading order for the coupling dependence
\begin{equation}
\Gamma_{1^+,1^-} \propto \left( \frac{g}{\omega_0} \right)^3.
\end{equation}
This result will account for the unexpected closure of the cascade channel.

\begin{figure}[t!]
	\centering
	\includegraphics[width=0.88\linewidth]{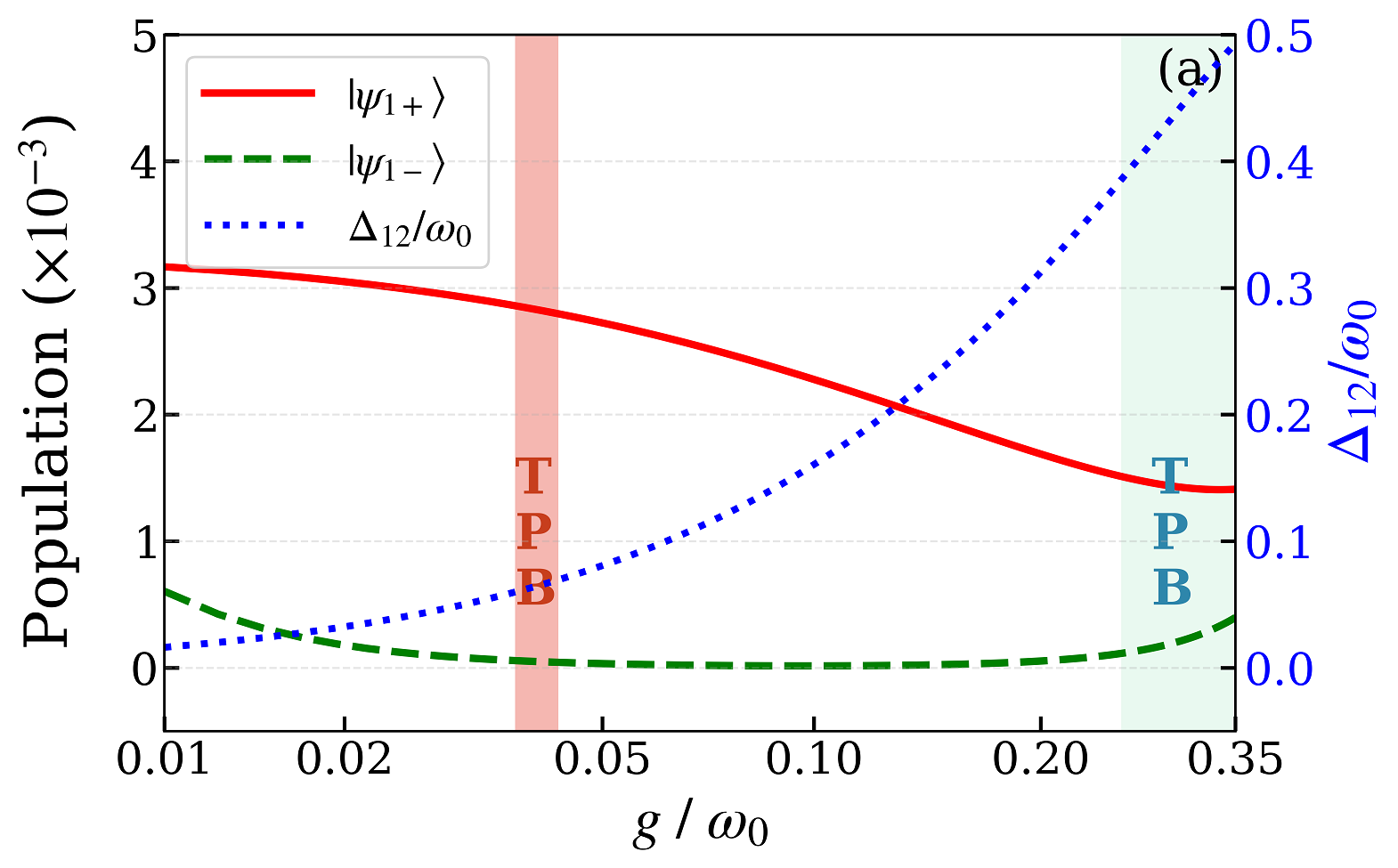}
	\includegraphics[width=0.88\linewidth]{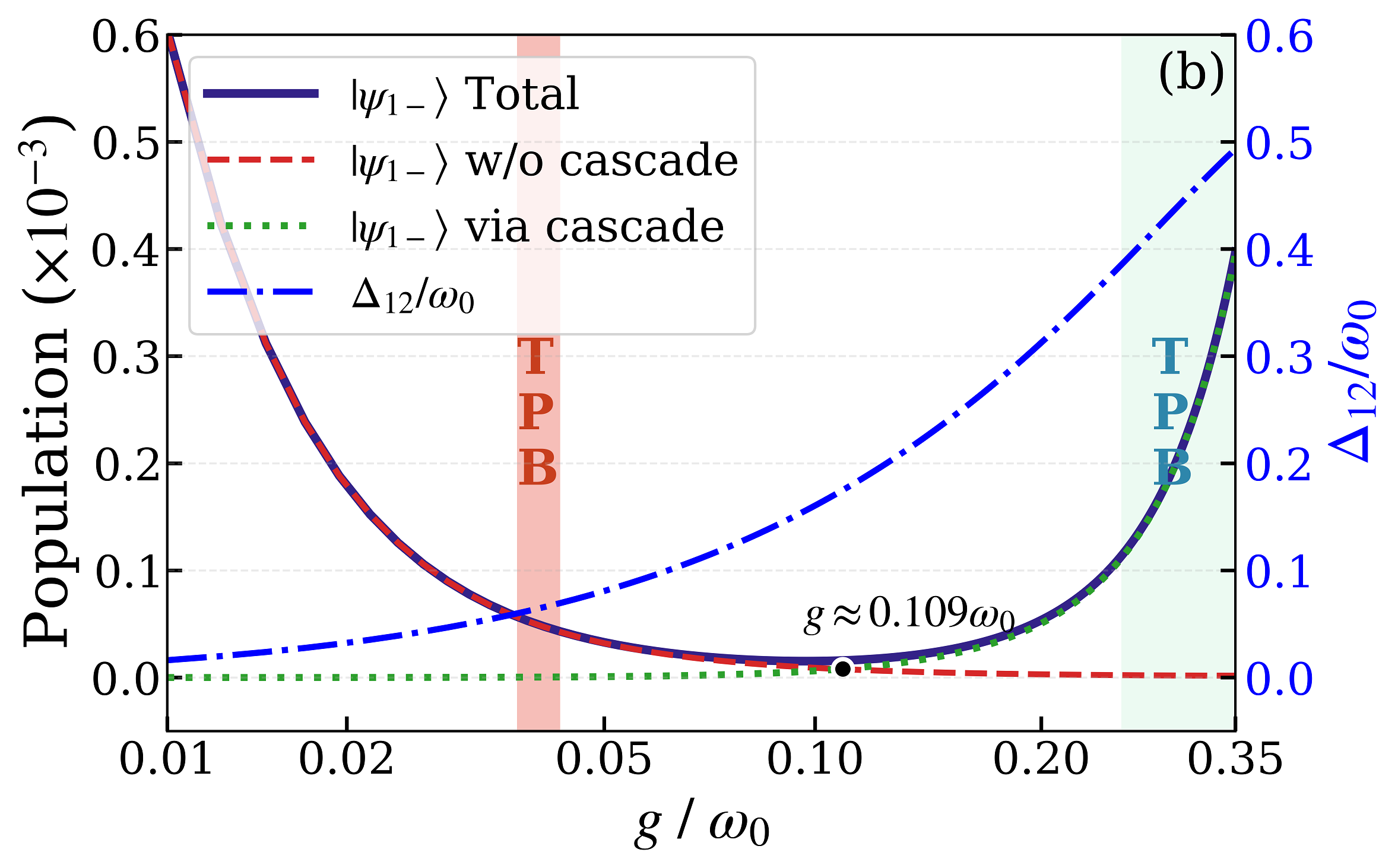}
	\caption{
{\it Different origins of population of $|\psi_{1-}\rangle$ in different TPBs.}
(a) Populations of different dressed states (left axis) and the energy splitting $\Delta_{12}$ (right axis) versus coupling strength $g$. Red solid, green dashed, and blue dotted lines represent the populations of $|\psi_{1+}\rangle$, $|\psi_{1-}\rangle$, and $\Delta_{12}$, respectively. In the weak TPB region, the $|\psi_{1-}\rangle$ population is small but nonzero.
(b) Contributions to the $|\psi_{1-}\rangle$ population from different mechanisms. Blue solid: total population; red dashed: direct driving; green dotted: cascade decay; blue dash-dotted: $\Delta_{12}$. The black dot marks the shift of two regimes: direct driving dominates for $g<0.109\omega_0$, while cascade decay dominates for $g>0.109\omega_0$.
}
\label{fig:pop}
\end{figure}

\subsection{Delayed opening of the cascade channel}

In Fig.~\ref{fig:annum} we compare the above perturbative results of the matrix elements $  |C_{1-,1+}^{(c)}| $ (dashed and dot-dashed lines), the energy splitting $\Delta_{12}$ (dotted line) and the final transition rate $\Gamma_{1+,1-} $ (solid line), with the exact numerical diagonalization (symbols). Here the prefactors in \eqref{eq-C-Matrix} are extracted to be $C_a(\theta) = 1.966$ and $C_{\sigma^-}(\theta)=(1+\sqrt{2})\cos(\theta)$. From the good agreements we see that the perturbative results really capture the leading-order behavior of the transition rate of the cascade channel.

From the verified perturbative expressions we can understand the unexpected closure of the cascade channel in the wide regime from the non-blockading phase, the weak-TPB phase to OPB phase. Indeed, the transition rate of the cascade channel is basically suppressed due to that the cubic law $\left( g/{\omega_0} \right)^3$ leads to a very slow growing of the transition rate with respect to small coupling, until significantly larger coupling strengths bring a quick increase in the regime where the coupling ratio $\left( g/{\omega_0} \right)$ being comparable to the unity. Thus, despite that the broken symmetry does not prohibit the cascade channel, it effectively appears to be close due to the very small amplitudes of transition rate. This cubic-law suppressing effect on the transition rate leads to a ``delayed" opening of the cascade channel until the transition to the strong-TPB phase.

\section{Physical mechanism for weak TPB: Role of weak ladder anharmonicity}\label{Sect-weak-TPB-mechanism}

A non-vanishing population of the $|\psi_{1-}\rangle  $ state is a key character of conventional TPB as in the strong TPB. In the OPB regime the $|\psi_{1-}\rangle$ population is vanishing, while the population becomes finite in the strong-TPB phase due to the opening of the cascade channel which allows the cascade decay from the $|\psi_{1+}\rangle$ state to the $|\psi_{1-}\rangle $ state. The importance of state population inspires us to track the population origin of the $|\psi_{1-}\rangle $ state as the starting clue, which finally redirects us to examine the population origin of the upper $|\psi_{2+}\rangle$ state as the right tracking direction. The origin excluding and clue redirecting lead us to reveal a different physical mechanism for the weak TPB.

\subsection{Residual population of $|\psi_{1-}\rangle  $}

To gain a clearer understanding of the physical mechanism underlying the weak-TPB phenomenon, we calculate the steady-state populations of different dressed states when the system is resonantly driven from $|\psi_0\rangle$ onto the $|\psi_{1+}\rangle  $ state. Indeed, as shown by the dashed line in Fig.~\ref{fig:pop}(a), the population of the $|\psi_{1-}\rangle  $ state is finite in the strong-TPB regime, while it is vanishing in the OPB phase, as expected. On the other hand, we notice that there is also some $|\psi_{1+}\rangle $ population in the weak-TPB window, although small compared with that of the $|\psi_{1+}\rangle  $ state, yet it is not zero.
This residual population originates from direct coherent driving.
For a more convincing analysis we will track the population by retaining and closing the dissipation channel in the following.

\subsection{Population of $|\psi_{1-}\rangle  $: Neither from cascade decay nor from the dephasing, but from driving}\label{Sect-popul-from-driving}

To distinguish the two contributions from cascade decay and driving, in Fig.~\ref{fig:pop}(b) we compare the $|\psi_{1-}\rangle  $ population obtained with the cascade channel artificially closed (red dashed line) and with the channel retained (green dotted line) respectively. As we can see, for $g < 0.109\omega_0 $, the population generated by direct driving dominates in the total population (purple solid line), whereas the cascade contribution becomes dominant for $g > 0.109\omega_0 $. The dot at $g\approx 0.109\omega_0$ marks the coupling point where the dominant contribution shifts from the driving to the cascade.

Note that the weak-TPB phase is located in the driving-contribution dominant regime where the cascade-decay contribution actually vanishes completely. These results demonstrate that the small $|\psi_{1-}\rangle  $ population observed in the weak-TPB regime arises from direct coherent driving rather than from the opening of the cascade channel. At this point, similar comparison also excludes the dephasing from key origin of the $|\psi_{1-}\rangle $ population, as addressed in Appendix~\ref{Appendix-TPB-on-dep}. The results also imply that the physical mechanism responsible for weak TPB differs fundamentally from that of the conventional TPB induced by activation of the cascade channel as in the strong TPB.

\begin{figure}[t!]
	\centering
	\includegraphics[width=0.8\linewidth]{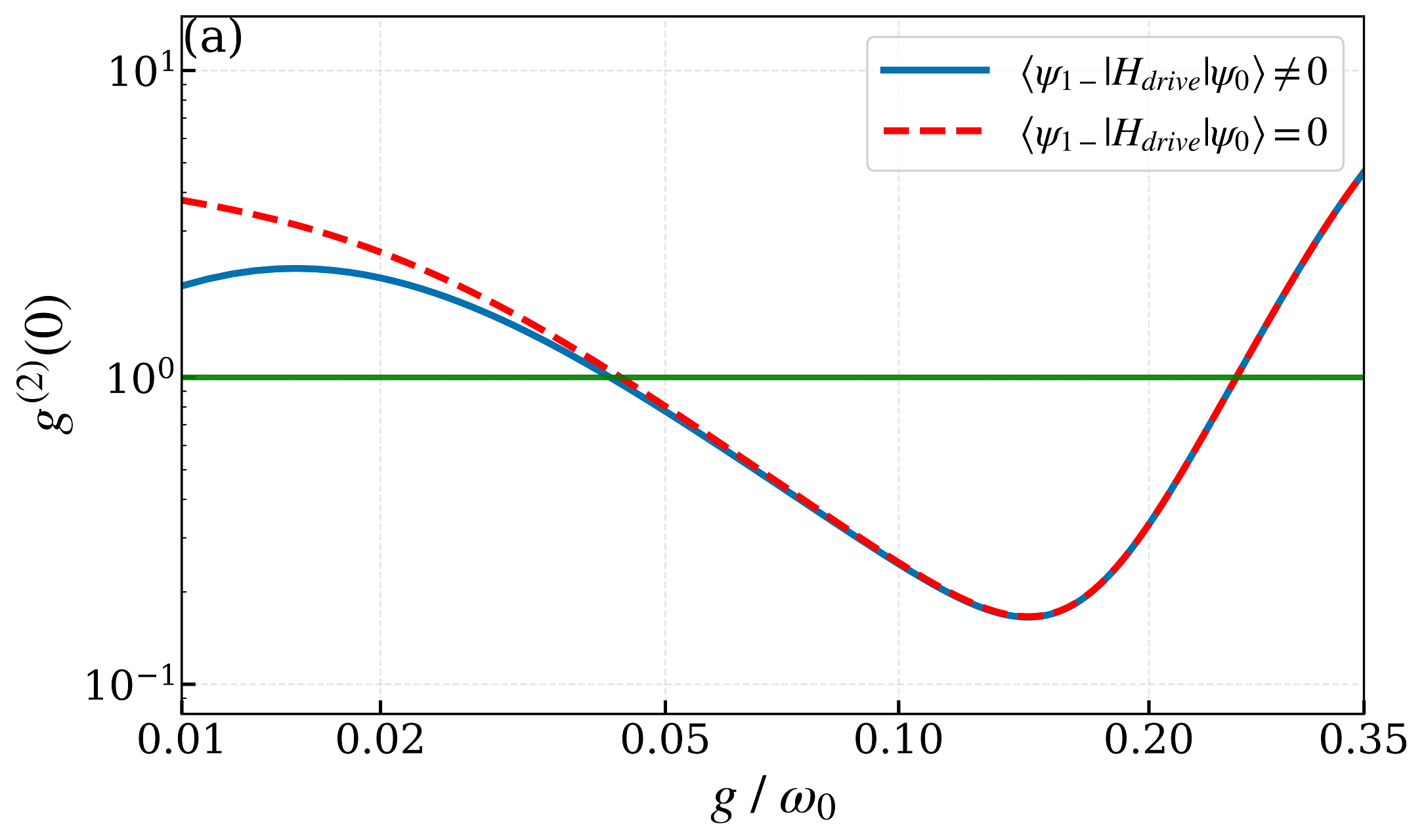}
	\includegraphics[width=0.8\linewidth]{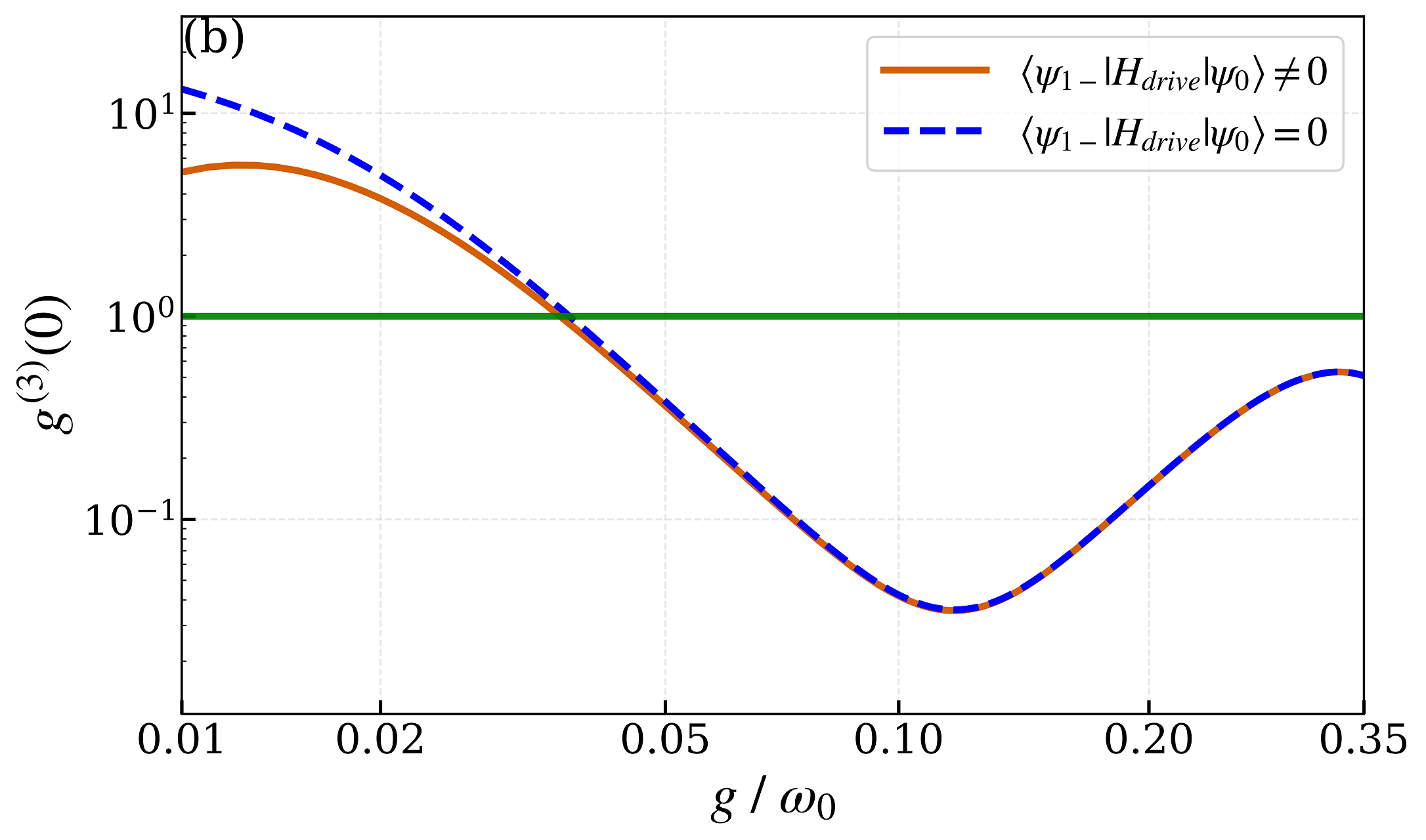}
	\caption{{\it Minor influence of driving-induced population of $|\psi_{1-}\rangle $ on $g^{(2)}(0)$ and $g^{(3)}(0)  $.}
(a) The second-order equal-time correlation function $g^{(2)}(0)$ as a function of the coupling strength $g$.
(b) The third-order equal-time correlation function $g^{(3)}(0)$ as a function of $g$.
In (a) and (b) the dashed lines correspond to the case where the driving matrix element to $|\psi_{1-}\rangle$ is artificially removed, while the red solid lines denote the results when it is retained.
}
\label{fig:g2gg}
\end{figure}

\subsection{Contrast to strong TPB: Minor influence of $|\psi_{1-}\rangle $ population on $g^{(2)}(0)  $ and $  g^{(3)}(0)  $}

In the above analysis we have found that the $|\psi_{1-}\rangle  $ population originates from the direct driving rather than the dissipation decay or dephasing. With the origin clarification, we can filter this origin artificially to analyze the actual influence on the second- and third-order correlation functions $g^{(2)}(0)  $ and $  g^{(3)}(0) $, as shown in Fig.~\ref{fig:g2gg}.
It turns out that, in the weak-TPB region, this direct-driving contribution only slightly reduces the values of both $g^{(2)}(0) $ [Fig.~\ref{fig:g2gg}(a)] and $g^{(3)}(0) $ [Fig.~\ref{fig:g2gg}(b)] and causes a minor shift in the position of the weak TPB window. To put it another way, even when the transition to $|\psi_{1-}\rangle  $ in direct driving is artificially turned off, both $g^{(2)}(0)$ and $g^{(3)}(0)$ still decrease below unity, and the weak TPB region persists without significant change. Thus, in a sharp contrast to the strong TPB where the $|\psi_{1-}\rangle $ population is a key character of mechanism for TPB (not in the driving as here but in the dissipation process), the influence of $|\psi_{1-}\rangle $ population is minor in the final formation of the weak TPB.

\begin{figure}[thbp]
    \includegraphics[width=0.88\linewidth]{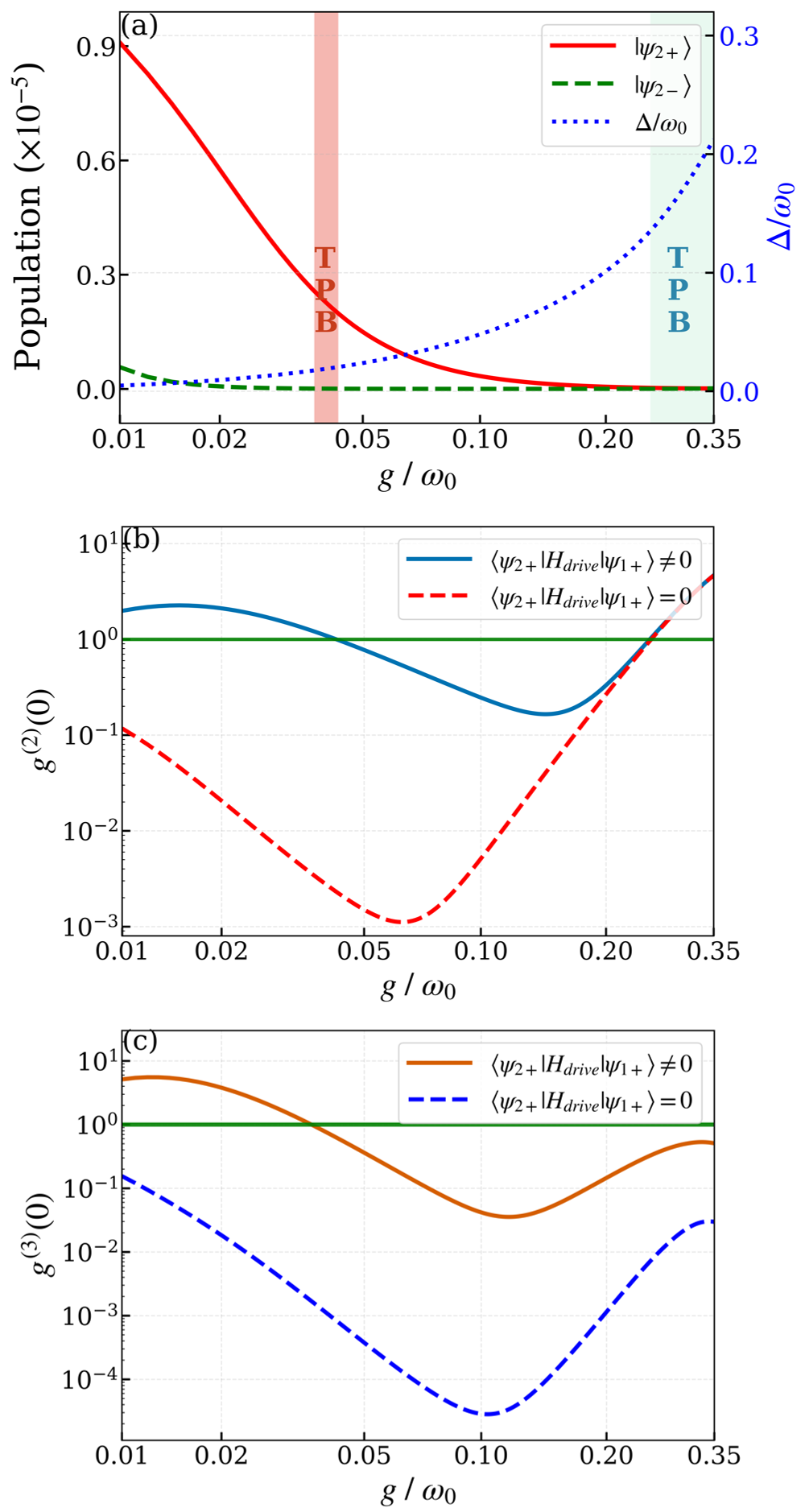}
	\caption{{\it Population of $|\psi_{2+}\rangle $ and crucial role of anharmonicity.}
(a) Populations of dressed states $|\psi_{2-}\rangle$ (dashed) and $|\psi_{2+}\rangle$ (solid) (left axis) and the anharmonicity parameter $\Delta$ (right axis) versus coupling strength $g$.
(b) The second-order equal-time correlation function $g^{(2)}(0)$ as a function of $g$.
(c) The third-order equal-time correlation function $g^{(3)}(0)$ as a function of $g$.
In (b)(c) the dashed lines correspond to the case where the driving matrix element to $|\psi_{2+}\rangle$ is artificially removed, while the red solid lines denote the results when it is retained.
}
\label{fig-anharmon}
\end{figure}

\subsection{Population of $|\psi_{2+}\rangle  $ and crucial role of weak anharmonicity}\label{Sect-role-anharmon}

In fact, the dominant physical mechanism underlying weak TPB is weak anharmonicity of the dressed-state ladder in the strong-coupling regime. The finding of the minor influence of $|\psi_{1-}\rangle $ population on the correlation functions in previous section redirects us to look into the population of upper states. Figure~\ref{fig-Ej} shows that, in the weak TPB region, the dressed energy levels in sector series $\{N,+\}$ nearly are equally spaced, which is a character of harmonicity. The harmonicity may induce some resonant driving to higher levels and open upper transition channel to accommodate photon processes, while anharmonicity will hinder such resonant driving.

Indeed, as shown by the dotted line in Fig.~\ref{fig-anharmon}(a), the anharmonicity parameter
\begin{equation}
\Delta = (E_{2^+} - E_{1+}) - (E_{1+} - E_0)
\end{equation}
is very small before the weak-TPB regime, indicating that the $  |\psi_0\rangle \to |\psi_{1^+}\rangle  $ and $  |\psi_{1^+}\rangle \to |\psi_{2^+}\rangle  $ transition frequencies are nearly identical. As a result, there are some population in the $|\psi_{2^+}\rangle  $ state, as shown by the solid line in Fig.~\ref{fig-anharmon}(a). This weak anharmonicity and finite $|\psi_{2^+}\rangle $ population keep both $  g^{(2)}(0)  $ and $  g^{(3)}(0)  $ greater than $1$ due to the open upper transition channel.

As the coupling strength $ g $ increases in Fig.~\ref{fig-anharmon}(a), the degree of anharmonicity grows, as indicated by the larger values of $\Delta$, which hinders the $|\psi_{2^+}\rangle $ population and causes both correlation functions to decrease. Because anharmonicity suppresses three-photon processes more strongly than two-photon processes, $  g^{(3)}(0)  $ falls below unity before $  g^{(2)}(0)  $ does, thereby forming the weak TPB region.  The dependence of the weak-TPB boundaries on the anharmonicity parameter in the presence of dissipation is given by Fig.~\ref{fig-gamma-Anharmonic} with a linear scaling relation in Appendix~\ref{Appendix-weak-TPB-vs-anharmon-gamma}.

To more definitely see the connection between the weak TPB and the weak-anharmonicity-induced population, in Fig.~\ref{fig-anharmon}(b) and~\ref{fig-anharmon}(c) we compare the correlation function $g^{(2)}(0)$ and $g^{(3)}(0)$ by keeping (solid lines) and turning off (dashed lines) the resonant transition from $|\psi_{1^+}\rangle$ to $|\psi_{2^+}\rangle$ in the driving. We see that turning off the transition to $|\psi_{2^+}\rangle$ dramatically reduces both the values of $g^{(2)}(0)$ and $g^{(3)}(0)$. Specially, the reduction of $g^{(2)}(0)$ far below unity completely destroys the weak TPB phase. As a matter of fact, both the non-blockading phase and the weak TPB phase are turned into OPB. In contrast, for the strong TPB $g^{(2)}(0)$ is not affected at all and the further reduction in $g^{(3)}(0)$ does not change the character as a TPB state. Such comparison demonstrates the crucial role of weak anharmonicity in the emerging of the weak-TPB phase.

\section{Comparison of the weak TPB and the strong TPB}\label{Sect-Compare}

To have a better view of the differences between the weak TPB and strong TPB, we summarize their main features in Table-\ref{Table-compare-2-TPB}.
The comparison shows that the weak TPB and strong TPB are essentially different in characters and key mechanisms despite that they both belong to TPB. Also we see that TPB does not necessarily originate from the cascade decay, while other reasons, as here the weak anharmonicity, also can lead to TPB. The different mechanism action stages also implies possible manipulation of TPB in different processes, as here the strong TPB in dissipation and the weak TPB in driving.

\begin{table}[t!]
\caption{Comparison of the weak TPB and the strong TPB }
\centering
\begin{tabular}{>{\raggedright}p{3.0cm}|>{\raggedright}p{2.6cm}|p{2.6cm}}
\hline\hline
Characters and mechanism                        & Strong TBP                   & Weak TPB\\[0.5ex]
\hline
Correlation functions                           & $g^{(2)}(0)>1 $ and $g^{(3)}(0)<1$     & $g^{(2)}(0)>1 $ and $g^{(3)}(0)<1$ \\
Coupling regime                                & Ultrastrong                            & Strong coupling \\
Key cascade decay channel~\footnotemark[1]     & Open                         & Close \\
Reason for cascade channel status              & Delayed by $g^3$ law         & Suppressed by $g^3$ law \\
Population on $|\psi_{1-}\rangle $             & Finite                       & Residual \\
Influence of $|\psi_{1-}\rangle $ population   & Crucial                      & Minor \\
Population on $|\psi_{2+}\rangle $             & Vanishingly small            & Finite \\
Influence of $|\psi_{2+}\rangle $ population   & Quantitative                 & Crucial \\
Location of key state                          & $|\psi_{1-}\rangle $, in the middle of cascade       & $|\psi_{2+}\rangle $, above the cascade states  \\
Primary mechanism                              & Cascade decay                & Weak anharmonicity  \\
Mechanism action stage                         & In dissipation               & In driving  \\
[1ex] \hline
\end{tabular}
\footnotetext[1]{The system is driven from the ground state $|\psi_0\rangle$ onto  $|\psi_{1+}\rangle $, the cascade states are $|\psi_{1+}\rangle \rightarrow |\psi_{1-}\rangle \rightarrow |\psi_0\rangle$ in dissipation. The key cascade channel for TPB is $|\psi_{1+}\rangle \rightarrow |\psi_{1-}\rangle$. }
\label{Table-compare-2-TPB}
\end{table}

\section{Conclusions}\label{Section-Conclusion}

In this work, we have uncovered a distinct weak TPB regime in a parity-broken generalized quantum Rabi model driven resonantly from the ground state onto the upper photon state $|\psi_{1+}\rangle$.
This weak TPB phase, manifesting the standard TPB characters of the second- and third-order correlation functions $g^{(2)}(0) > 1  $ and $g^{(3)}(0) < 1  $, emerges within a window in the strong-coupling regime, well separated from the strong TPB in the USC regime. While the strong TPB occurs conventionally with opening of the cascade decay channel from $|\psi_{1+}\rangle$ to $ |\psi_{1-}\rangle $ denoted by the decay rate $\Gamma_{1+,1-} $, the weak TPB appears unconventionally with the cascade channel closed.

In fact, the closure of the cascade decay channel in the wide coupling regime before the strong TPB is unexpected in the sense that the presence of the parity symmetry breaking should have allowed the opening of the cascade channel. To reveal the mechanism underlying this delayed opening of cascade channel, by means of perturbative analysis in the dressed-state basis, we demonstrated that the suppression of the cascade channel originates from destructive quantum interference. Indeed,
with the broken parity symmetry, the transition matrix elements $\langle\psi_{1-}|(c-c^\dagger)|\psi_{1+}\rangle$ contribute to the cascade decay rate $\Gamma_{1+,1-} $ with a $\mathcal{O}[(g/\omega_0)^2] $ term. With the addition of the mainly linear dependence in the energy splitting, the cascade decay rate finally is governed by a cubic law $\mathcal{O}[(g/\omega_0)^3] $ in the leading order. Consequently, with the slow growing rate in the beginning of the cubic law, the cascade rate is effectively suppressed in the wide regimes of weak coupling, strong coupling and even some range of USC, until the transition of the strong TPB around the coupling ratio $g/\omega_0\approx 0.253$ quite deep into the USC regime. The cubic law of the cascade decay rate accounts for the delayed opening of the cascade channel, whereas it also leaves the enigma of the weak TPB unconventionally with closed cascade channel for us to resolve.

To find out the underlying mechanisms for the weak TPB, we have tracked the origin of the population of upper states.  In the strong TPB, the vanishing or finite population of the $|\psi_{1-}\rangle $ state is the result and also the fingerprint of the closing or opening of the cascade channel. We found that in the weak TPB the $|\psi_{1-}\rangle $ population is not vanishing either, which however is not coming from cascade decay dissipation or depahsing, but from resonant driving. Still, judging from the minor the influence on the correlation functions $g^{(2)}(0)$ and $g^{(3)}(0)$ we concluded that the $|\psi_{1-}\rangle $ population does not play the role for the occurrence of the weak TPB. We finally found that it is weak anharmonicity of the energy-level ladder that plays the essential role. In fact, the weak anharmonicity leads to a non-vanishing population of higher state $|\psi_{2+}\rangle $ above the cascade levels in the resonant driving which opens the new channel to influence the photon blockade. Indeed, by artificially turning off the resonant driving to the $|\psi_{2+}\rangle $ state, the correlation functions are qualitatively changed and the weak TPB is destroyed. Thus, we have clarified that, while the cascade decay in dissipation leads to the strong TPB, weak anharmonicity in driving is responsible for the weak TPB.

Our analyses may provide insights for the generation of photon blockades which has promising application potential for advanced quantum technologies on the manipulation level of individual quanta. As an implication, our results illustrate that multi-photon blockade in strongly coupled light-matter systems can arise through multiple, physically distinct pathways. The weak TPB identified here relies on the interplay between parity-breaking-induced interference (which keeps the cascade channel closed) and weak anharmonicity (which differentially affects two- and three-photon processes), rather than on real population transfer via cascade decay. Moreover, our mechanism tracking also indicates that the manipulation of TPB can be intervened in different stages, as we see here in driving process for weak TBP and in dissipation process for strong TBP. The mechanism and manipulation diversities offer more opportunities for the design of non-classical photon sources and quantum gates in circuit-QED platforms operating at different coupling strengths, where both counter-rotating terms and engineered parity breaking can be controllably exploited.

\section*{Acknowledgments}
This work was supported by the National Natural Science Foundation of China (Grants No. 12474358 and No. 12247101).

\appendix

\section{Dependence of the weak-TPB phase on the anharmonicity and dissipation parameters with linear boundary scaling}\label{Appendix-weak-TPB-vs-anharmon-gamma}

\begin{figure}[thbp]
	\centering
	\includegraphics[width=0.93\linewidth]{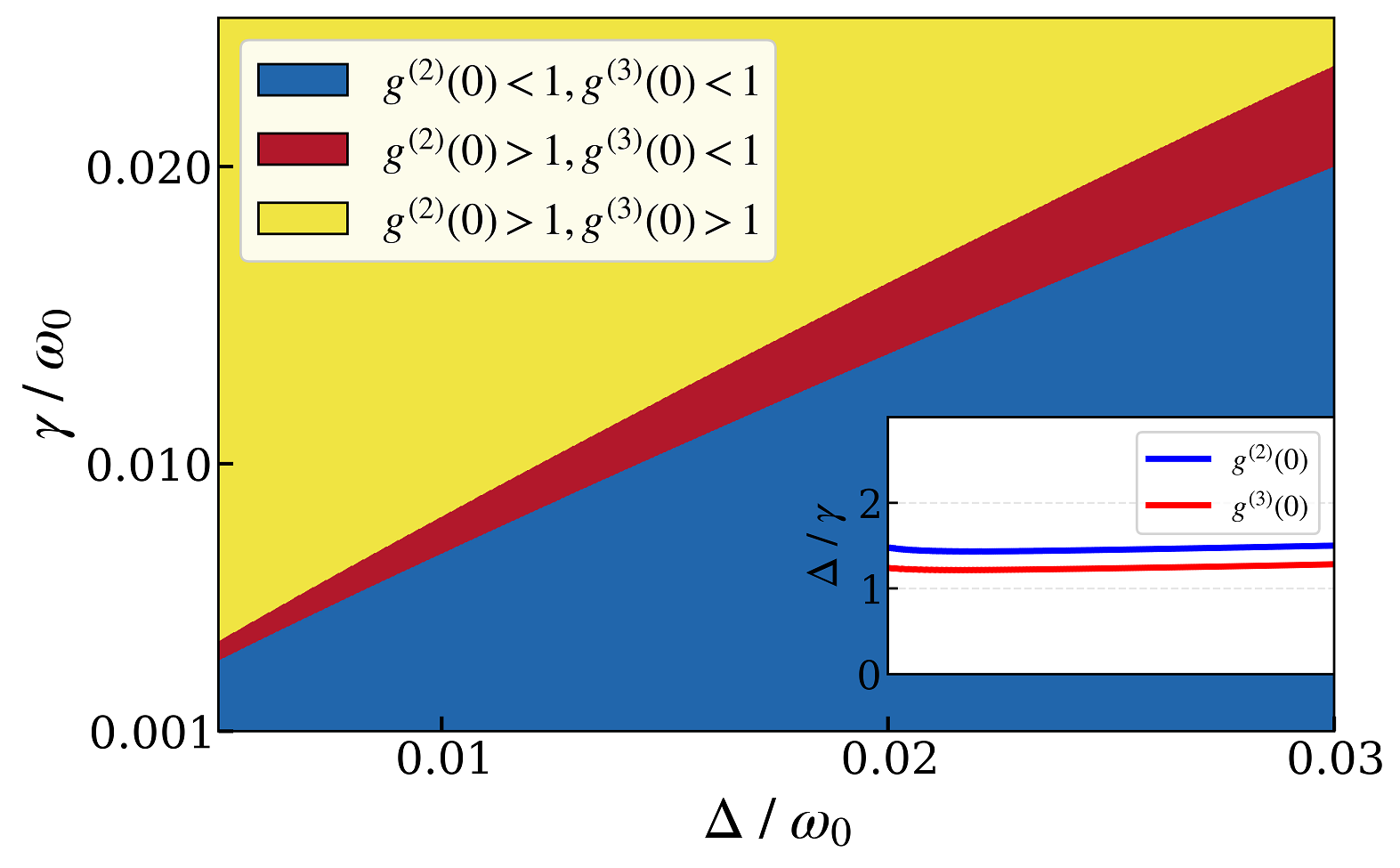}
	\caption{{\it Dependence of the weak-TPB phase on the anharmonicity parameter $\Delta$ and dissipation parameter $\gamma$.} The phase diagram in the main panel is divided into the non-blockading phase [yellow (light gray) region] with $g^{(2)}(0) >1$ and $g^{(3)}(0)>1$, the OPB phase [blue (gray) region] with $g^{(2)}(0) <1$ and $g^{(3)}(0)<1$, and the weak-TPB phase [red (dark gray)] with $g^{(2)}(0) >1$ and $g^{(3)}(0)<1$. The Inset shows the ratio $\Delta/\gamma$ at the lower [red (dark gray)] and upper [blue (gray)] boundaries of the weak-TPB phase with $g^{(3)}(0)=1$ and $g^{(2)}(0)=1$ respectively. The horizontal axis and ticks (unlabeled) of the Inset are the same as the main panel.}
\label{fig-gamma-Anharmonic}
\end{figure}

\begin{figure}[h!]
	\centering
	\includegraphics[width=0.9\linewidth]{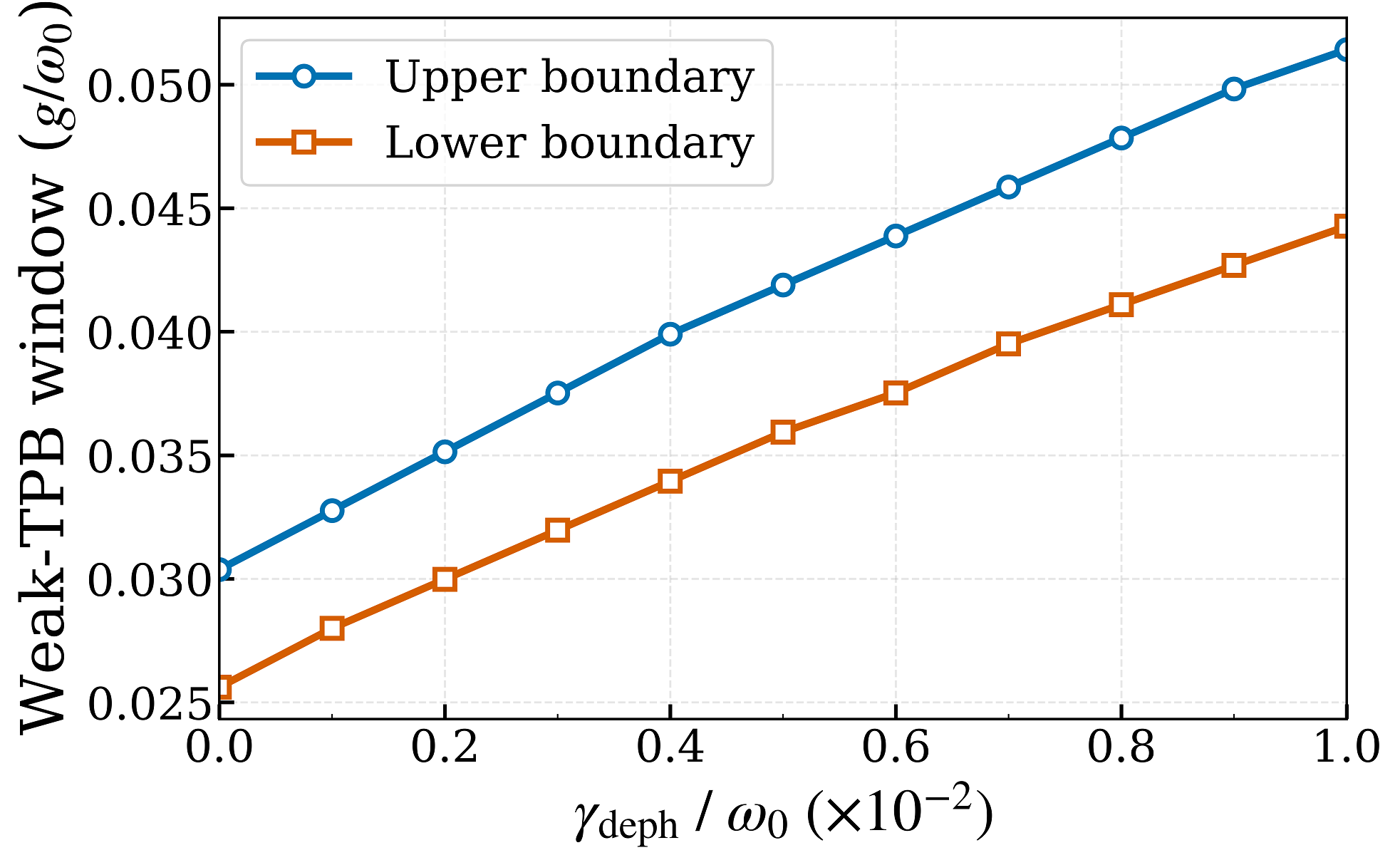}
	\caption{Dependence of the lower (squares) and upper (circles) boundaries of the weak-TPB phase in the coupling $g$ dimension on the qubit dephasing rate $\gamma_{\rm deph}$. Here, the dissipation parameters $\gamma_a = \gamma_{\sigma^-} = 0.01\omega_0$ are held fixed.
}
\label{fig-weak-TPB-gc-vs-deph}
\end{figure}

In Sec.~\ref{Sect-role-anharmon}, we have figured out that weak anharmonicity is the primary mechanism for and the formation of the weak-TPB phase. In the main text we illustrated the weak-TPB in the presence of dissipation with a fixed dissipation parameters $\gamma=\gamma_a = \gamma_{\sigma^-}=10^{-2}\omega_0  $. Although the key cascade channel $|\psi_{1+}\rangle \rightarrow |\psi_{1-}\rangle$, which makes the mechanism difference for the weak TPB and strong TPB, is closed in this regime, there is a non-vanishing population on the $|\psi_{1-}\rangle$ state due to driving as tracked in Sec.~\ref{Sect-popul-from-driving}. Furthermore, the other dissipation channels $|\psi_{1+}\rangle \rightarrow |\psi_{0}\rangle$ and $|\psi_{1-}\rangle \rightarrow |\psi_{0}\rangle$ are still open, as shown by the finite transition rates $\Gamma _{1+,0}$ and $\Gamma _{1-,0}$ in Fig.~\ref{fig:g2g3}. So the dissipation will still influence the weak-TPB boundaries indirectly. Here, we provide a phase diagram for the weak-TPB in the plane of the weak anharmonicity parameter $\Delta$ and the dissipation parameter $\gamma$, as shown in Fig.~\ref{fig-gamma-Anharmonic}. The yellow (light gray) region is the non-blockading phase, as denoted by $g^{(2)}(0) >1$ and $g^{(3)}(0)>1$. The blue (gray) region represents the OPB phase where $g^{(2)}(0) <1$ and $g^{(3)}(0)<1$. The red (dark gray) window between the non-blockading phase and the OPB phase is the weak-TPB phase with $g^{(2)}(0) >1$ and $g^{(3)}(0)<1$. Here the anharmonicity parameter $\Delta$ is tuned by varying the coupling strength $g$. As we see, with the increase of $\gamma$ the window of the anharmonicity opened for the weak TPB becomes wider. It turns out that the weak TPB boundaries manifest a linear variation in the typically used regime of $\gamma$ shown in Fig.~\ref{fig-gamma-Anharmonic}. Indeed, as more explicitly displayed in the Inset of Fig.~\ref{fig-gamma-Anharmonic}, the values of $\Delta$ and $\gamma$ scale with each other by nearly constant ratios,
\begin{equation}
  \frac{\Delta}{\gamma} \simeq {\mathrm constant},
\end{equation}
at the lower and upper boundaries of the weak TPB phase.

\section{Dependence of TPB boundries on dephasing}\label{Appendix-TPB-on-dep}

As a fuller consideration of the environmental situation, actually in Fig.~\ref{fig:pop}(c) we have also included the qubit dephasing by a finite value of $\gamma _{\rm deph}$. Nevertheless, the dephasing is not the essential origin of the weak TPB. In fact, we have also tracked the origin of the $|\psi_{1-}\rangle  $ population by excluding the dephasing, in the same way as in Fig.~\ref{fig:pop}(c). It turns out that excluding and including dephasing only make a quantitative difference for the position of weak TPB phase, while the population contributions qualitatively remain the same as Fig.~\ref{fig:pop}(c). Indeed, we also examine the variation of the upper and lower boundaries of the weak-TPB phase when the $\gamma _{\rm deph}$ is changed, as shown in Fig.~\ref{fig-weak-TPB-gc-vs-deph}. We see that the weak-TPB boundaries shift to higher values of coupling when the $\gamma _{\rm deph}$ is increasing,  however the weak-TPB phase is still there when the dephasing is turned off at $\gamma _{\rm deph}=0$. This result indicates that essentially the $|\psi_{1-}\rangle  $ population of $|\psi_{1-}\rangle  $ and the emerging of the weak TPB are neither originating from the dephasing.

\section{Perturbative proof that the dissipation matrix element $M$ scales
as $g^{2}$}

\label{Appendix-proof-M-g2}

This appendix presents a rigorous perturbative derivation for the
contribution of the dissipation matrix element
\begin{equation}
\left\vert C_{1-,1+}^{(c)}\right\vert =\bigl|\langle \psi
_{1-}|(c-c^{\dagger })|\psi _{1+}\rangle \bigr|,
\end{equation}%
where $c=a$ or $c=\sigma ^{-}$, in the transition rate in dissipation. This
quantity controls the cascade decay rate and exhibits distinct leading
behaviors in the scaling of the coupling strength $g$: For the cavity
dissipation channel ($c=a$), we find $|C_{1-,1+}^{(a)}|\propto g^{2}$,
implying $|C_{1-,1+}^{(a)}|^{2}\propto g^{4}$; For the qubic dissipation
channel ($c=\sigma ^{-}$), the scaling is linear $|C_{1-,1+}^{(\sigma
^{-})}|\propto g$, so that $|C_{1-,1+}^{(\sigma ^{-})}|^{2}\propto g^{2}$.
The proof is carried out exclusively through explicit algebraic summation
and term-by-term sign tracking in the $N=0$ and $N=2$ level sectors. All
formulas below are derived self-consistently from the definitions in Sec.~%
\ref{sec:setup}.

\subsection{Perturbation treatment and definitions}

\label{sec:setup} We work in dimensionless unit by setting $\omega _{0}=1$.
The Hamiltonian is divided into the Jaynes--Cummings (JC) part $H_{\mathrm{JC%
}}$ and non-JC part $V$. The JC Hamiltonian is taken as the unperturbed part
which reads
\begin{equation}
H_{\mathrm{JC}}=N+g\beta \,(a\sigma ^{+}+a^{\dagger }\sigma ^{-}),
\end{equation}%
where $N=a^{\dagger }a+\sigma _{z}/2+1/2$ is the excitation number and $%
\beta =\sin \theta $, while the non-JC part
\begin{equation}
V=g\alpha \,(a+a^{\dagger })\sigma _{z}-g\beta \,(a\sigma ^{-}+a^{\dagger
}\sigma ^{+}),
\end{equation}%
with $\alpha =\cos \theta $, is treated as the perturbation. The unperturbed
states and energies are analytically available and explicitly given by
\begin{eqnarray}
&&|\psi _{N\pm }^{\left( 0\right) }\rangle =\frac{1}{\sqrt{2}}(|e,N-1\rangle
\pm |g,N\rangle ),  \label{state-expression-N} \\
&&E_{N\pm }^{(0)}=N\pm \sqrt{N}g\beta
\end{eqnarray}%
for $N\geqslant 1$, while $|\psi _{0}\rangle ^{(0)}=|g,0\rangle $ and $%
E_{0}^{(0)}=0$ for the ground state with $N=0$. Here $g,e$ represent the
ground (spin-down) state and excited (spin-up) state of the qubit, while $N$
is the photon number in the Fock basis and equals to the excitation number.

\subsection{Vanishing zeroth order}\label{Appendix-zero-order}

According to the output-field definition in Eq.~\eqref{Eq-input-ouput}, the
transition operator is
\begin{equation}
X^{c}=c-c^{\dagger }\quad (c=a,\sigma ^{-}),
\end{equation}%
whose matrix element \eqref{Eq-Cjk} contributes to the decay rate in %
\eqref{eq:t10}. In the zeroth order, the matrix element vanishes exactly
\begin{equation}
\left( C_{1-,1+}^{(c)}\right) _{00}=\left\langle \psi _{1-}\left\vert
X^{c}\right\vert \psi _{1+}\right\rangle ^{(0)}=0,  \label{Eq-C11-0th}
\end{equation}%
where the superscript $\left( 0\right) $ denotes the expectation over the
zeroth-order wave function, due to that $X^c$ breaks the symmetry of
excitation number and induces transition from the $\psi _{1+}$ state in the $%
N=1$ sector to states in other $N$ sectors which are orthogonal to the $N=1$
sector.

\subsection{First-order correction in perturbation}

\label{Appendix-Perturb-Order-1}

The first-order corrections to the wave functions in bra and ket are
\begin{eqnarray}
\langle \psi _{1-}^{(1)}| &=&\sum_{v\neq \psi _{1+},\psi _{1-}}\frac{\langle
\psi _{1-}|V|v\rangle ^{(0)}}{E_{1-}^{(0)}-E_{v}^{(0)}}\langle v|^{(0)},
\label{Eq-waveF-correction-1A} \\
|\psi _{1+}^{(1)}\rangle  &=&\sum_{v\neq \psi _{1+},\psi _{1-}}|v\rangle
^{(0)}\frac{\langle v|V|\psi _{1+}\rangle ^{(0)}}{E_{1+}^{(0)}-E_{v}^{(0)}},
\label{Eq-waveF-correction-1B}
\end{eqnarray}%
Note that the states $v=\psi _{1,\pm }^{^{(0)}}$ do not contribute in %
\eqref{Eq-waveF-correction-1A} and \eqref{Eq-waveF-correction-1B} also due
to vanishing matrix element in Eq. \eqref{Eq-C11-0th}. Substituting the
corrections \eqref{Eq-waveF-correction-1A} and \eqref{Eq-waveF-correction-1B}
into the dissipative matrix element yields the first-order correction

\begin{equation}
C_{1-,1+}^{\left( c\right) }=\left( C_{1-,1+}^{\left( c\right) }\right)
_{00}+\left( C_{1-,1+}^{\left( c\right) }\right) _{01}+\left(
C_{1-,1+}^{\left( c\right) }\right) _{11}
\end{equation}%
where%
\begin{eqnarray}
\left( C_{1-,1+}^{\left( c\right) }\right) _{00} &=&\langle \psi
_{1-}^{(0)}|X^{c}|\psi _{1+}^{(0)}\rangle  \\
\left( C_{1-,1+}^{\left( c\right) }\right) _{11} &=&\langle \psi
_{1-}^{(1)}|X^{c}|\psi _{1+}^{(1)}\rangle  \\
\left( C_{1-,1+}^{\left( c\right) }\right) _{01} &=&\langle \psi
_{1-}^{(0)}|X^{c}|\psi _{1+}^{(1)}\rangle +\langle \psi
_{1-}^{(1)}|X^{c}|\psi _{1+}^{(0)}\rangle
\end{eqnarray}%
are the zeroth order, the bilinear term and the mixed term, respectively.
The zeroth order vanishes as in Eq.\eqref{Eq-C11-0th}, the bilinear term is
of higher order than the mixed term. The mixed term can be written more
explicitly
\begin{equation}
\begin{split}
\left( C_{1-,1+}^{\left( c\right) }\right) _{01}=\sum_{v}\Biggl[& \frac{%
\langle \psi _{1-}|X^{c}|v\rangle ^{(0)}\times \langle v|V|\psi _{1+}\rangle
^{(0)}}{E_{1+}^{(0)}-E_{v}^{(0)}} \\
& +\frac{\langle \psi _{1-}|V|v\rangle ^{\left( 0\right) }\times \langle
v|X^{c}|\psi _{1+}\rangle ^{\left( 0\right) }}{E_{1-}^{(0)}-E_{v}^{(0)}}%
\Biggr].
\end{split}
\label{Eq-Cc1}
\end{equation}

To extract the leading orders, we can carry out expansion on the mixed term
\begin{equation}
\left( C_{1-,1+}^{\left( c\right) }\right) _{01}=S+P+\mathcal{O}(g^{3})
\end{equation}%
with respect to the coupling $g$ by expanding the denominators in the
summation
\begin{equation}
\begin{split}
\frac{1}{E_{1+}^{(0)}-E_{v}^{(0)}}& =\frac{1}{D_{v}}-\frac{g\beta }{D_{v}^{2}%
}+\mathcal{O}(g^{2}), \\
\frac{1}{E_{1-}^{(0)}-E_{v}^{(0)}}& =\frac{1}{D_{v}}+\frac{g\beta }{D_{v}^{2}%
}+\mathcal{O}(g^{2}).
\end{split}
\label{Eq-denominnators-expand}
\end{equation}%
where $D_{v}=E_{c}-E_{v}^{(0)}$ and $E_{c}=1$. Here $S$ and $P$ correspond to the zeroth and linear term in the denominator expansion
\begin{equation}
\begin{split}
S=\sum_{v}\Bigl[\langle \psi _{1-}|& X^{c}|v\rangle ^{\left( 0\right)
}\langle v|V|\psi _{1+}\rangle ^{\left( 0\right) } \\
& +\langle \psi _{1-}|V|v\rangle ^{\left( 0\right) }\langle v|X^{c}|\psi
_{1+}\rangle ^{\left( 0\right) }\Bigr]\frac{1}{D_{v}}
\end{split}
\label{Eq-Cc1-S}
\end{equation}%
and%
\begin{equation}
\begin{split}
P=\sum_{v}\Bigl[-\langle \psi _{1-}|& X^{c}|v\rangle ^{\left( 0\right)
}\langle v|V|\psi _{1+}\rangle ^{\left( 0\right) } \\
& +\langle \psi _{1-}|V|v\rangle ^{\left( 0\right) }\langle v|X^{c}|\psi
_{1+}\rangle ^{\left( 0\right) }\Bigr]\frac{g\beta }{D_{v}^{2}}.
\end{split}
\label{Eq-Cc1-P}
\end{equation}%
The contribution $S$ still contains both linear ($J$) and quadratic ($K_S$) terms
\begin{equation}
S = J + K_S +\mathcal{O}(g^{3}),
\end{equation}
while $P$ does not have any linear term. We can collect all quadratic terms from $S$ and $P$
\begin{equation}
K = K_S +K_P,
\end{equation}
so that the mixed matrix element can be decomposed into the linear term $J$ and the quadratic term $K$,
\begin{equation}
\left( C_{1-,1+}^{\left( c\right) }\right) _{01}= J + K + \mathcal{O}(g^{3}) .
\end{equation}%
Then, we can finally find out the leading order for $c=a,\sigma ^{-}$, respectively, by carrying out the summation $v$ over
the $N=0$ and $N=2$ sectors as in the following, while the other sectors do not contribute as $X^{c}$ only induces transitions between bases in neighbor sectors.

\subsection{Proof that the linear-order term in $C_{1-,1+}^{(a)}$ vanishes}

We first consider the $c=a$ case. We will prove $J=0$ and $K\neq 0$ in this cavity dissipation channel by checking
the relevant $N=0$ and $N=2$ sectors in the summation over $v$.

\subsubsection{$N=0$ sector}

Take the virtual state $v=|\psi _{0}^{(0)}\rangle =|g,0\rangle $ (with $%
D_{0}= 1$). We evaluate the two contributions in Eq.~\eqref{Eq-Cc1-S}
separately.

For the first term in Eq.~\eqref{Eq-Cc1-S}, we have $X^{a}|\psi
_{0}^{(0)}\rangle =-|g,1\rangle $, which immediately implies $\langle \psi
_{1-}|X^{a}|\psi _{0}\rangle ^{(0)}=+1/\sqrt{2}$. Applying the perturbation
gives
\begin{equation}
V|\psi _{1+}^{(0)}\rangle =\frac{1}{\sqrt{2}}g\alpha \left[ |e,1\rangle
-|g,0\rangle -\sqrt{2}|g,2\rangle \right] -g\beta |e,2\rangle ,  \label{V-psi-1+}
\end{equation}%
so that $\langle \psi _{0}|V|\psi _{1+}\rangle ^{(0)}=-g\alpha /\sqrt{2}$.
The product of the two matrix elements in the first term of Eq.~%
\eqref{Eq-Cc1-S} is therefore $(+1/\sqrt{2})\times (-g\alpha /\sqrt{2}%
)=-g\alpha /2$. Dividing by the energy denominator $D_{0}$
produces the contribution $-g\alpha /2$.

For the second term in Eq.~\eqref{Eq-Cc1-S}, the action of the perturbation
on the other state yields
\begin{equation}
V|\psi _{1-}^{(0)}\rangle =\frac{1}{\sqrt{2}}g\alpha \left[ |e,1\rangle
+|g,0\rangle +\sqrt{2}|g,2\rangle \right] +g\beta |e,2\rangle ,  \label{V-psi-1-}
\end{equation}%
from which it follows that $\langle \psi _{0}|V|\psi _{1-}\rangle
^{(0)}=+g\alpha /\sqrt{2}$. The lowering component of $X^{a}$ gives $\langle
\psi _{0}|X^{a}|\psi _{1+}\rangle ^{(0)}=+1/\sqrt{2}$. The resulting product
is $(+g\alpha /\sqrt{2})\times (+1/\sqrt{2})=+g\alpha /2$, and division by $D_{0}$ produces the contribution $+g\alpha /2$.

Adding the two contributions yields exact cancellation
\begin{equation}
-g\alpha /2+g\alpha /2=0
\end{equation}%
for the $N=0$ sector in contribution to $S$.

\subsubsection{$N=2$ sector}

The states in the $N=2$ sector are composed of bases $|g,2\rangle $ and $%
|e,1\rangle$ as in Eq.~\eqref{state-expression-N}.  For the contribution involving $|g,2\rangle $, we have
\begin{equation}
\langle \psi _{1-}^{(0)}|X^{a}|g,2\rangle =-1\quad \text{and}\quad \langle
g,2|V|\psi _{1+}^{(0)}\rangle =-g\alpha ,
\end{equation}%
which immediately gives
\begin{equation}
\langle \psi _{1-}^{(0)}|X^{a}|g,2\rangle \times \langle g,2|V|\psi
_{1+}^{(0)}\rangle =g\alpha
\end{equation}%
as the contribution to the first term of $S$. The order-transpose element
from the second term in $S$ is
\begin{equation}
\langle \psi _{1-}^{(0)}|V|g,2\rangle =g\alpha \quad \text{and}\quad \langle
g,2|X^{a}|\psi _{1+}^{(0)}\rangle =-1,
\end{equation}%
yielding
\begin{equation}
\langle \psi _{1-}^{(0)}|V|g,2\rangle \times \langle g,2|X^{a}|\psi
_{1+}^{(0)}\rangle =-g\alpha .
\end{equation}%
Adding these two contributions leads to exact cancellation
\begin{equation}
g\alpha -g\alpha =0.
\end{equation}%
in $S$ for the basis $|g,2\rangle $

Similarly, for the contribution involving $|e,1\rangle $, the matrix
elements satisfy
\begin{equation}
\langle \psi _{1-}^{(0)}|X^{a}|e,1\rangle =\frac{1}{\sqrt{2}}\quad \text{and}%
\quad \langle e,1|V|\psi _{1+}^{(0)}\rangle =\frac{g\alpha }{\sqrt{2}},
\end{equation}%
so that
\begin{equation}
\langle \psi _{1-}^{(0)}|X^{a}|e,1\rangle \times \langle e,1|V|\psi
_{1+}^{(0)}\rangle =\frac{g\alpha }{2}.
\end{equation}%
The matrix elements in the second term of $S$ reads
\begin{equation}
\langle \psi _{1-}^{(0)}|V|e,1\rangle =\frac{g\alpha }{\sqrt{2}}\quad \text{%
and}\quad \langle e,1|X^{a}|\psi _{1+}^{(0)}\rangle =-\frac{1}{\sqrt{2}},
\end{equation}%
giving
\begin{equation}
\langle \psi _{1-}^{(0)}|V|e,1\rangle \times \langle e,1|X^{a}|\psi
_{1+}^{(0)}\rangle =-\frac{g\alpha }{2}.
\end{equation}%
These two contributions likewise cancel exactly:
\begin{equation}
\frac{g\alpha }{2}-\frac{g\alpha }{2}=0.
\end{equation}

It should be noted that there exist mixed-basis terms, $\langle \psi_{1-}^{(0)} | V | g,2 \rangle \times \langle e,1 | X^{a} | \psi_{1+}^{(0)} \rangle = -g\alpha/\sqrt{2}  $ and $  \langle \psi_{1-} | X^{a} | e,1 \rangle^{(0)} \langle g,2 | V | \psi_{1+} \rangle^{(0)} = -g\alpha/\sqrt{2}  $, which are equal and cannot cancel out. Nevertheless, because the sign in front of the basis state $  |g,2\rangle  $ is opposite in the two unperturbed states $  |\psi_{2+}^{(0)}\rangle  $ and $  |\psi_{2-}^{(0)}\rangle  $ [see Eq.~\eqref{state-expression-N}], these mixed terms produce contributions of order $  g^{2}  $ when summed over $  |\psi_{2+}^{(0)}\rangle  $ and $  |\psi_{2-}^{(0)}\rangle  $.
Indeed, although these mixed terms themselves contribute at order $g$, the energy denominators $  D_{v}  $ associated with $  |\psi_{2+}^{(0)}\rangle  $ and $  |\psi_{2-}^{(0)}\rangle  $ are
$D_{2+}=  -1 - \sqrt{2} g \beta  $
and
$D_{2-}=  -1 + \sqrt{2} g \beta  $,
respectively. Their difference
\begin{equation}
\frac{1}{-1 - \sqrt{2} g \beta} - \frac{1}{-1 + \sqrt{2} g \beta} = \frac{2\sqrt{2} g \beta}{1 - 2 g^{2} \beta^{2}} \label{Dv-Dv-N=2}
\end{equation}
produces a linear factor in Eq.~\eqref{Eq-Cc1-S}. Consequently, the mixed terms contribute to $K_S$ at order $g^{2} $ after the summation.

Since both $|\psi _{2+}^{\left( 0\right) }\rangle $ and $|\psi _{2-}^{\left(
0\right) }\rangle $ are a linear combination of $|e,1\rangle $ and $%
|g,2\rangle $, their contributions at $g$ order completely cancel in Eq.~\eqref{Eq-Cc1-S}
according to the above separate-basis analysis. Consequently, we find
\begin{equation}
J=0\ \text{for }c=a.
\end{equation}%
In other words, the linear-order part of $C_{1-,1+}^{(a)}$ vanishes exactly.

\subsubsection{Origins of the quadratic term}

Since both the zeroth-order and the linear term vanishes, the leading
non-zero contribution in $C_{1-,1+}^{(a)}$ is the quadratic term. The
quadratic term arises from four sources that follow directly from the
expansion in Sec.~\ref{Appendix-Perturb-Order-1}:

\begin{enumerate}

\item The uncanceled mixed-basis terms in $S$ in the $N=2$ sector with the $D_v$ difference in Eq.~\eqref{Dv-Dv-N=2}, which contributes the $g^2$-order term $K_S$.

\item The second-order piece $P$ generated by expanding the energy
denominators in Eq.~\eqref{Eq-denominnators-expand}, which contributes the $g^2$-order term $K_P$.

\item The bilinear term
\begin{equation}
\begin{split}
& \langle \psi _{1-}|X^{a}|\psi _{1+}\rangle ^{(1)} \\
& =\sum_{v,w}\frac{\langle \psi _{1-}|V|v\rangle ^{(0)}\ \langle
v|X^{a}|w\rangle ^{(0)}\ \langle w|V|\psi _{1+}\rangle ^{(0)}}{%
(E_{1-}^{(0)}-E_{v}^{(0)})(E_{1+}^{(0)}-E_{w}^{(0)})}
\end{split}%
\end{equation}%
from inclusion of both corrections from the bra and ket parts in Eqs.%
\eqref{Eq-waveF-correction-1A} and \eqref{Eq-waveF-correction-1B}.

\item The second-order wave-function corrections
\begin{equation}
\langle \psi _{1-}^{(0)}|X^{a}|\psi _{1+}^{(2)}\rangle +\langle \psi
_{1-}^{\left( 2\right) }|X^{a}|\psi _{1+}^{0}\rangle .
\end{equation}
\end{enumerate}

Collecting these contributions yields
\begin{equation}
|C_{1-,1+}^{(a)}|=C_{a}(\theta )\left( \frac{g}{\omega _{0}}\right) ^{2}+%
\mathcal{O}(g^{3}),
\end{equation}%
where the dimensionless coefficient $C(\theta )\approx 1.966$ at $\theta
=0.3\pi $ which is consistent with the result from numerical diagonalization.

\subsection{Proof for non-vanishing linear term in $C_{1-,1+}^{(\protect\sigma ^{-})}$}

Now we check $c=\sigma ^{-}$ case. We prove $J\neq 0$ by explicit
summation over the relevant $N=0$ and $N=2$ sectors.

\subsubsection{$N=0$ sector}

For the contribution involving $|\psi _{0}\rangle =|g,0\rangle $, the matrix
elements satisfy
\begin{equation}
\langle \psi _{1-}|X^{\sigma ^{-}}|\psi _{0}\rangle ^{(0)}=-\frac{1}{\sqrt{2}%
}\quad \text{and}\quad \langle \psi _{0}|V|\psi _{1+}\rangle ^{(0)}=-\frac{%
g\alpha }{\sqrt{2}},
\end{equation}%
so that
\begin{equation}
\langle \psi _{1-}|X^{\sigma ^{-}}|\psi _{0}\rangle ^{(0)}\times \langle
\psi _{0}|V|\psi _{1+}\rangle ^{(0)}=\frac{g\alpha }{2}.
\label{Spin-matrix-1}
\end{equation}%
The sector-transpose elements from the second sum in Eqs.~\eqref{Eq-Cc1-S} and \eqref{Eq-Cc1-P} read
\begin{equation}
\langle \psi _{1-}|V|\psi _{0}\rangle ^{(0)}=\frac{g\alpha }{\sqrt{2}}\quad
\text{and}\quad \langle \psi _{0}|X^{\sigma ^{-}}|\psi _{1+}\rangle ^{(0)}=%
\frac{1}{\sqrt{2}},
\end{equation}%
giving
\begin{equation}
\langle \psi _{1-}|V|\psi _{0}\rangle ^{(0)}\times \langle \psi
_{0}|X^{\sigma ^{-}}|\psi _{1+}\rangle ^{(0)}=\frac{g\alpha }{2}.
\label{Spin-matrix-2}
\end{equation}%
The two contributions from \eqref{Spin-matrix-1} and \eqref{Spin-matrix-2}
act oppositely in $S$ and $P$
\begin{eqnarray}
\frac{g\alpha }{2}+\frac{g\alpha }{2} &=&g\alpha \quad \text{in}\quad S, \\
-\frac{g\alpha }{2}+\frac{g\alpha }{2} &=&0\quad \text{in}\quad P.
\end{eqnarray}%
Thus, the $N=0$ sector has a finite contribution to $S\ $but cancelled
contrubition to $P$.

\subsubsection{$N=2$ sector}

For the contribution involving the $|e,1\rangle $ basis in the $N=2$ sector
of states, the matrix elements satisfy
\begin{equation}
\langle \psi _{1-}^{(0)}|X^{\sigma ^{-}}|e,1\rangle =-\frac{1}{\sqrt{2}}%
\quad \text{and}\quad \langle e,1|V|\psi _{1+}^{(0)}\rangle =\frac{g\alpha }{%
\sqrt{2}},
\end{equation}%
so that
\begin{equation}
\langle \psi _{1-}^{(0)}|X^{\sigma ^{-}}|e,1\rangle \times \langle
e,1|V|\psi _{1+}^{(0)}\rangle =-\frac{g\alpha }{2}. \label{Spin-matrix-3}
\end{equation}%
The sector-transpose elements from the second sum in Eqs.~\eqref{Eq-Cc1-S} and \eqref{Eq-Cc1-P} reads
\begin{equation}
\langle \psi _{1-}^{(0)}|V|e,1\rangle =\frac{g\alpha }{\sqrt{2}}\quad \text{%
and}\quad \langle e,1|X^{\sigma ^{-}}|\psi _{1+}^{(0)}\rangle =-\frac{1}{%
\sqrt{2}},
\end{equation}%
giving
\begin{equation}
\langle \psi _{1-}^{(0)}|V|e,1\rangle \times \langle e,1|X^{\sigma
^{-}}|\psi _{1+}^{(0)}\rangle =-\frac{g\alpha }{2}. \label{Spin-matrix-4}
\end{equation}%
These two contributions from \eqref{Spin-matrix-2} and \eqref{Spin-matrix-2} yield a non-vanishing linear term in $S$ despite that they cancel in $P$:
\begin{eqnarray}
-\frac{g\alpha }{2}-\frac{g\alpha }{2} &=&-g\alpha \quad \text{in}\quad S, \\
+\frac{g\alpha }{2}-\frac{g\alpha }{2} &=&0\quad \text{in}\quad P.
\end{eqnarray}%
For the contribution involving the $|g,2\rangle $ basis in the $N=2$ sector
of states, the matrix elements satisfy $\langle \psi _{1-}^{(0)}|X^{\sigma ^{-}}|g,2\rangle =0$ and
$\langle g,2|X^{\sigma ^{-}}|\psi _{1+}^{(0)}\rangle =0$. Therefore, in the $N=2$ sector, the $|g,2\rangle$ basis has no contribution.

For the mixed-basis term in the case $  c = \sigma^-  $, the analysis follows a procedure analogous to that for $  c = a  $. The difference lies in the matrix element of the mixed-basis term,
\begin{equation}
\langle \psi_{1-}^{(0)} | X^{\sigma^-} | e,1 \rangle = -\frac{1}{\sqrt{2}},
\end{equation}%
which is opposite in sign to the $  c = a  $ case,
$\langle \psi_{1-}^{(0)} | X^{a} | e,1 \rangle = +\frac{1}{\sqrt{2}}$, while $\langle e,1|X^{c}|\psi _{1+}^{(0)}\rangle =-\frac{1}{\sqrt{2}}$ are the same for $c = a,\sigma^-$.
Therefore,
\begin{equation}
\langle \psi_{1-}^{(0)} | V | g,2 \rangle \times \langle e,1 | X^{\sigma^-} | \psi_{1+}^{(0)} \rangle = -\frac{g\alpha}{\sqrt{2}},
\end{equation}
and
\begin{equation}
\langle \psi_{1-}^{(0)} | X^{\sigma^-} | e,1 \rangle \langle g,2 | V | \psi_{1+}^{(0)} \rangle = +\frac{g\alpha}{\sqrt{2}}.
\end{equation}
have opposite signs and already cancel each other in the same state of $v$ before the summation in the $N=2$ sector, unlike the $  c = a  $ case in Eq.~\eqref{Dv-Dv-N=2}.

Consequently, the complete summation over the $N=0$ and $N=2$ sectors proves
\begin{eqnarray}
S&&=g\alpha\frac{1}{D_0}+(\frac{1}{D_{2+}}+\frac{1}{D_{2+}})(-\frac{1}{\sqrt{2}}g\alpha) \nonumber\\
 &&=(1+\frac{\sqrt{2}}{1-2g^2\beta^2})g\alpha \nonumber\\
 &&\approx (1+\sqrt{2})g\alpha
\end{eqnarray}
and $P=0$. Thus, we have finite linear term of $%
|C_{1-,1+}^{(\sigma ^{-})}|$.

\section{Perturbative expansion of the energy splitting $\Delta_{12} = E_{1+} -
E_{1-}$}
\label{Appendix-gap}

The energy splitting $\Delta_{12} =E_{1+}-E_{1-}$ between the dressed states
$|\psi _{1+}\rangle $ and $|\psi _{1-}\rangle $ also admits a systematic
perturbative expansion in powers of $g$. The unperturbed value (from the JC
part $H_{\mathrm{JC}}$) is
\begin{equation}
\Delta_{12} ^{(0)}=E_{1+}^{(0)}-E_{1-}^{(0)}=2g\beta .
\end{equation}%
The first-order correction vanishes identically,
\begin{equation}
\Delta_{12} ^{(1)}=\langle \psi _{1\pm }^{(0)}|V|\psi _{1\pm }^{(0)}\rangle
=0,
\end{equation}%
as $V$ induces transitions to bases of neighbor sectors ($N=0,1$) or
next-neighbor sector ($N=3$) which are orthogonal to the $N=1$ bases. Thus
the first-order correction to $\Delta_{12} $ is zero.

The second-order correction to each individual energy
\begin{equation}
E_{1\pm }^{(2)}=\sum_{v\neq \psi _{1+},\psi _{1-}}\frac{|\langle v|V|\psi
_{1\pm }\rangle ^{(0)}|^{2}}{E_{1\pm }^{(0)}-E_{v}^{(0)}}.
\end{equation}%
sums up the contributions of virtual states excluding the $N=1$ sector,
finally from $N=0,2,3$ sectors.

More explicitly, for $N=0$ the unerturbative energy is $E_{v}^{(0)}=0$ while
$E_{1\pm }^{(0)}=1\pm g\beta $, and from Eqs. (\ref{V-psi-1+}) and (\ref%
{V-psi-1-}) we have $\langle \psi _{0}|V|\psi _{1\pm }\rangle ^{(0)}=\mp
g\alpha /\sqrt{2}$. Thus%
\begin{eqnarray}
\Delta _{12,N=0}^{(2)} &=&E_{1+,N=0}^{(2)}-E_{1-,N=0}^{(2)}  \notag \\
&=&\frac{|-g\alpha /\sqrt{2}|^{2}}{1+g\beta }-\frac{|+g\alpha /\sqrt{2}|^{2}%
}{1-g\beta }  \notag \\
&=&-g^{3}\alpha ^{2}\beta +\mathcal{O}(g^{5}).
\end{eqnarray}

For the $N=2$ sector, we have the energy difference in the denominator%
\begin{eqnarray}
E_{1\pm }^{(0)}-E_{2+}^{(0)} &=&-1+\left( -\sqrt{2}\pm 1\right) g\beta , \\
E_{1\pm }^{(0)}-E_{2-}^{(0)} &=&-1+\left( \sqrt{2}\pm 1\right) g\beta ,
\end{eqnarray}%
and%
\begin{eqnarray}
\left\vert \langle \psi _{2+}|V|\psi _{1\pm }\rangle ^{(0)}\right\vert ^{2}
&=&\frac{1}{4}g^{2}\alpha ^{2}\left( 1\mp \sqrt{2}\right) ^{2}, \\
\left\vert \langle \psi _{2-}|V|\psi _{1\pm }\rangle ^{(0)}\right\vert ^{2}
&=&\frac{1}{4}g^{2}\alpha ^{2}\left( 1\pm \sqrt{2}\right) ^{2}.
\end{eqnarray}%
Then we get%
\begin{eqnarray}
\Delta _{12,N=2}^{(2)} &=&\frac{-7g^{3}\alpha ^{2}\beta -2g^{4}\beta ^{4}}{%
1-6g^{2}\beta ^{2}+g^{4}\beta ^{4}}  \notag \\
&=&-7g^{3}\alpha ^{2}\beta +\mathcal{O}(g^{4}).
\end{eqnarray}

For the $N=3$ sector, we have the energy difference in the denominator%
\begin{eqnarray}
E_{1\pm }^{(0)}-E_{3+}^{(0)} &=&-2+\left( -\sqrt{3}\pm 1\right) g\beta , \\
E_{1\pm }^{(0)}-E_{3-}^{(0)} &=&-2+\left( \sqrt{3}\pm 1\right) g\beta ,
\end{eqnarray}%
and%
\begin{equation}
\left\vert \langle \psi _{3+}|V|\psi _{1\pm }\rangle ^{(0)}\right\vert
^{2}=\left\vert \langle \psi _{3-}|V|\psi _{1\pm }\rangle ^{(0)}\right\vert
^{2}=\frac{1}{2}g^{2}\beta ^{2}.
\end{equation}%
So we find%
\begin{eqnarray}
\Delta _{12,N=3}^{(2)} &=&\frac{-g^{3}\beta ^{3}\left( 2+g^{2}\beta
^{2}\right) }{4-8g^{2}\beta ^{2}+g^{4}\beta ^{4}}  \notag \\
&=&-g^{3}\beta ^{3}/2+\mathcal{O}(g^{5}).
\end{eqnarray}%
The second-order correction is then summed up to be
\begin{eqnarray}
\Delta _{12}^{(2)} &=&\Delta _{12,N=0}^{(2)}+\Delta _{12,N=2}^{(2)}+\Delta
_{12,N=3}^{(2)}  \notag \\
&=&-(8\alpha ^{2}\beta +\beta ^{3}/2)g^{3}+\mathcal{O}(g^{4})
\end{eqnarray}%
in the leading order.

Collecting all these orders yields
\begin{equation}
\Delta _{12}=2g\beta -(8\alpha ^{2}\beta +\beta ^{3}/2)g^{3}+\mathcal{O}%
(g^{5}).
\end{equation}%
In the weak TPB regime, which is within the range $g/\omega _{0}\lesssim
0.043$, the $\mathcal{O}(g^{3})$ correction is negligible, so that the
linear leading order $\Delta _{12}\approx 2g\beta $ has a high accuracy.

\bibliography{Refs-2026-6-2Photon-Blockade}

\end{document}